\documentclass[numberedappendix,appendixfloats,iop]{emulateapj}
\usepackage{natbib}
\usepackage[utf8x]{inputenc}
\usepackage{amssymb}
\usepackage{amsmath}
\usepackage{graphicx}
\usepackage{setspace}
\usepackage{color}
\newcommand{\dd}{\text{d}}

\def \bo {\text{bo}}
\def \dyn {\text{dyn}}
\def \diff {\text{diff}}
\def \lum {\text{ls}}

\def \BB {\text{BB}}
\def \cl {\text{cl}}
\def \s {\text{s}}

\def \dyn {\text{dyn}}
\def \diff {\text{diff}}
\def \obs {\text{obs}}

\def \d {\text{d}}
\def \bol {\text{bol}}
\def \min {\text{min}}
\def \max {\text{max}}
\def \sa {\text{sa}}
\def \ff {\text{ff}}

\def \ej {\text{ej}}

\def \pl {\text{pl}}

\def \tr {\text{tr}}
\def \e {\text{c}}
\def \c {\text{c}}
\def \eq {\text{eq}}
\def \p {\text{pl}}

\begin{document}

\title{Early supernova emission - logarithmic corrections to the planar phase}
\author{Tamar Faran and Re'em Sari}
\affiliation{Racah Institute of Physics, The Hebrew University, Jerusalem 91904, Israel}

\email{tamar.faran@mail.huji.ac.il}
\begin{abstract}
When the shock wave generated in a supernova explosion breaks out of the stellar envelope, the first photons, typically in the X-ray to UV range, escape to the observer. Following this breakout emission, radiation from deeper shells diffuses out of the envelope as the supernova ejecta expands. Previous studies have shown that the radiation throughout the planar phase (i.e., before the expanding envelope has doubled its radius) originates in the same mass coordinate, called the `breakout shell'. We derive a self-similar solution for the radiation inside the envelope, and show that this claim is incorrect, and that the diffusion wave propagates logarithmically into the envelope (in Lagrangian sense) rather than remaining at a fixed mass coordinate. The logarithmic correction implies that the luminosity originates in regions where the density is $\sim 10$ times higher than previously thought, where the photon production rate is increased and helps thermalization. We show that this result has significant implications on the observed temperature. In our model, the radiation emitted from blue supergiant and Wolf-Rayet explosions is still expected to be out of thermal equilibrium during the entire planar phase, but the observed temperature will decrease by two orders of magnitude, contrary to previous estimates. Considering the conditions at the end of the planar phase, we also find how the temperature and luminosity transition into the spherical phase.
%Consequently, the observed radiation will peak at lower frequencies - far UV for Blue Supergiants and soft X-rays for Wolf-Rayet stars.
\end{abstract}

\section{Introduction}
The first emission from a supernova (SN) explosion is called the `shock breakout', and marks the emergence of a shock wave from the stellar envelope. During the explosion, a radiation dominated shock wave accelerates in the decreasing density profile of the progenitor star. It deposits its energy into the envelope, leaving behind it an expanding radiation dominated gas. Once the width of the shock becomes comparable to the distance from the edge of the star, the photons that power the shock wave diffuse ahead of it and the shock "breaks out" of the envelope \citep[]{colgate74,falk78,klein_chevalier78,imshennik_etal81,ensman_borrows92,matzner&mckee}. This breakout emission typically lasts for a few seconds, and is followed by a long phase of photon diffusion, which produces emission that decays slowly \citep[]{ns10,rabinak11,Piro2010}.
%The light emitted from the supernova can be powered either by the energy deposited by the shock wave (shock cooling radiation), or by radioactive decay of unstable isotopes that were synthesized in the explosion, e.g. $^{56}$Ni. The shock cooling emission is dominant in massive progenitors that have retained an extensive gas envelope, such as type-II SNe. These explosions do not produce a large amount of radioactive material (typically a few 0.01Msun, cite), but since their envelopes are initially very extended, not all of the shock energy is lost to adiabatic expansion. Consequently, the dominant contribution to the luminsoity comes from the energy deposited by the shock. However, in explosions from very compact progenitors, such as type Ia SNe, most of the energy goes into the expansion of the envelope, resulting in a very dim shock cooling radiation, never to have been detected in observations. Nevertheless, a type Ia explosion synthesizes a large amount of radioactive elements (a few $\sim 0.1 M_{\odot}$), and the observed light curves are therefore very bright due to radioactive decay ($\sim 10^{43}$ erg). 
%
%The two energy components can dominate the SN luminosity at different times. The light curves of Type-II SNe typically show a long phase which is dominated by shock cooling emission. After all the energy from the shock has been radiated away, a sharp drop in the luminosity is observed and the SN enters the radioactive tail phase. During this phase, the decay of $^{56}$Co into $^{56}$Fe is the dominant process that governs the light curve.

Immediately after breakout, the evolution of the ejecta is described by planar expansion, during which the radius of the envelope remains roughly constant. The duration of the planar phase depends on the radius of the progenitor and on the shock velocity, and typically lasts between a few minutes in compact progenitors to hours for more extended stars. After the radius of the SN has doubled its size, the envelope enters the spherical phase, in which the ejecta expands homologously. In this paper we focus on the planar phase evolution. This phase was investigated in very few works, including \cite{ns10} who examined the post breakout emission for various types of progenitors, and \cite{Piro2010} who looked at the breakout from type Ia SNe. Both of these works assumed that the diffusion wave stays at a fixed mass, implying that the same material is the source of radiation throughout the entire planar phase.
They estimated that the breakout emission peaks somewhere between the UV and soft $\gamma$ rays, depending on the type of the progenitor. As the SN ejecta expands and cools adiabatically, the radiation is shifted to lower frequencies.
Early shock breakout emission might have been observed as an X-ray outburst in the two cases of SN2006aj and SN2008D \citep{campana06,Soderberg2008b}. Later optical emission suggested that both were Ib/c SNe, whose progenitors are compact Wolf-Rayet (WR) stars \citep{pian06,mazzali06,modjaz06,maeda07,mazzali08,malesani09}.
%and at the end of the planar phase, it is in the optical for Red Supergiants (RSGs), and UV for Blue Supergiants (BSGs).

In this work we find self-similar solutions for the radiation in the SN envelope.
We first solve for the hydrodynamic profiles of the SN envelope after shock passage and during the planar expansion, according to \cite{sakurai60}. We then use these profiles as initial conditions for the following stage of photon diffusion. This derivation takes into account the density and shock velocity profiles (given that they can be described as polytropes), and is therefore expected to produce more accurate solutions than the approximate order of magnitude models.
A self-similar solution for the radiation of type-II SNe was also found by \cite{Chevalier1992}. However, they treated only the spherical phase, and did not provide a solution for the earlier planar phase.

The most important result of this paper is that through the similarity solution, we were able to find a logarithmic correction to the dynamics of the planar phase, compared to what was thought so far. We find the mass from which radiation escapes to the observer at the end of the planar phase to be an order of magnitude larger than estimated before. Correspondingly, the emission peaks at lower frequencies than would have been inferred without the logarithmic correction.

In Section \ref{s:pre_explosion_hydro} we describe the pre-explosion structure of the progenitor and in Section \ref{s:planar_hydro} we find the planar hydrodynamic evolution of the envelope. We treat the breakout phase and find a self-similar solution for the energy in the envelope including diffusion in Section \ref{s:diffusion}. We then provide the resulting luminosity in Section \ref{s:luminosity} and observed temperature in Sections \ref{s:T_obs}. We apply the results of this paper to various progenitors in Section \ref{s:progenitors}.

\section{Pre-expansion hydrodynamics} \label{s:pre_explosion_hydro}
The radiation-dominated shock wave that forms during the explosion propagates through the ejecta, and accelerates in the decreasing density profile of the envelope. The pre-explosion density profile of the progenitor star near its surface, can be approximated as a power-law of the distance from the edge of the star, $R-r$, namely 
\begin{equation} \label{e:rho}
\rho\propto (R-r)^n ~,
\end{equation}
where $R$ is the stellar radius, $r$ is the distance from the center of the star and $n=3/2$ for radiative envelopes while $n=3$ for convective envelopes. This density profile is a simplification that describes well only the outer regions of the stellar envelope. Since in this work we focus on the first seconds to hours of the SN evolution, the above description for the density profile serves as a good approximation.

The hydrodynamics of shock passage through a stellar envelope in the form of Eq \eqref{e:rho} were treated in \cite{sakurai60} (hereafter S60). As in S60, we assume that the shock wave is a discontinuity that reaches all the way to the edge of the star, i.e. $R-r=0$. In reality, the shock has a certain width, and it breaks out of the envelope when its width is comparable to the distance to the edge of the star. However, an accurate description of the dynamics at breakout lies beyond the scope of this paper. We therefore adopt the simplistic view of S60, but note that our analysis may not describe well material lying ahead of the breakout location.

Throughout this work, we follow the notations of S60 for the pre-explosion profile of the star. The density is parametrized as
\begin{equation}
\label{e:rho_i}
\rho_i = \kappa_1 (R-r)^n,
\end{equation}
and the shock velocity is
\begin{equation}
\label{e:u_shock}
U_{shock} = \kappa_2 (R-r)^{-\mu n},
\end{equation}
where $\mu$ is determined by requiring continuity at the sonic point, while $\kappa_1$ and $\kappa_2$ are constants that depend on the progenitor mass $M$, radius $R$ and explosion energy $E$ of the progenitor star. We choose the following representation for $\kappa_1$ and $\kappa_2$:
\begin{equation} \label{e:kappa}
\begin{split}
&\kappa_1 = C_1 \frac{M}{R^{n+3}}\\
&\kappa_2 = C_2 \bigg(\frac{E}{M}\bigg)^{1/2} R^{\mu n} ~.
\end{split}
\end{equation}
The constants $C_1$ and $C_2$ are dimensionless constants, that depend on the inner structure of the progenitor star. In Section \ref{s:progenitors} we estimate their values for different progenitors.

Using equations \eqref{e:rho_i} and \eqref{e:u_shock} we find self-similar solutions for the hydrodynamic equations  (equations 2--4 in S60). We do not provide the full calculation here, but note that the only difference from S60's work is that we do not consider an external gravitational force, since the energy of the explosion is typically much larger than the gravitational energy, i.e. $GM^2/R\ll E$.
For given values of $n$ and the adiabatic index $\gamma$ there is a single value of $\mu$ that satisfies the continuity condition at the sonic point. Since the value of $\mu$ is insensitive to $n$, we use the canonical value of $\mu=0.19$.
For further details on the solution, we refer the reader to S60.
%For $n=3/2$ and $\gamma=4/3$ we find $\mu = 0.191$ and $\mu=0.186$ for $n=3$. 
After the passage of the shock, the ejecta gains kinetic energy and starts to expand, while decreasing in density and internal energy. The following hydrodynamic evolution can be divided into two temporal regimes - a planar phase, during which the radius of the envelope is nearly constant ($r\simeq R$), and a spherical phase that starts after the envelope has roughly doubled its size, where the radius evolves homologously and satisfies $r=v\times t$. In this paper we treat only the planar phase and the short transition into the spherical phase, while taking into account both adiabatic cooling and photon diffusion.
%The conditions at the end of the planar phase serve as initial conditions for the spherical phase which will be addressed in a following paper.

\section{Planar hydrosynamics} \label{s:planar_hydro}
The stage of expansion into vacuum was first treated in S60. We repeat the procedure and solve the hydrodynamic equations (equations 16--19 in S60) using the solution of the preceding stage when the shock is at $r=R$ as initial conditions. Here instead of the spatial Lagrangian coordinate `$a$' used in S60, we write the expressions in terms of the Lagrangian mass coordinate `m', which is related to $a$ by:
\begin{equation}\label{e:m}
m = \int 4\pi R^2 \rho_0 ~da~,
\end{equation}
where $\rho_0$ is the density profile in the envelope when the shock reaches the stellar edge.
We find the self-similar solutions for the fluid velocity $v$, density $\rho$, pressure $P$ and the distance from the pre-expansion stellar radius $x$, during the planar expansion phase:
\begin{subequations} \label{e:planar_selfsim_hydro}
\begin{align}
&x(m,\xi) = \widetilde{x} \Big(\frac{m}{M}\Big)^{1/(n+1)} r(\xi) \\
\label{e:v_xi}
&v(m,\xi) = \widetilde{v}\Big(\frac{m}{M}\Big)^{-\mu n/(n+1)} F(\xi) \\
&\rho(m,\xi) = \widetilde{\rho}\Big(\frac{m}{M}\Big)^{n/(n+1)}H(\xi) \\
&P(m,\xi) =  \widetilde{P} \Big(\frac{m}{M}\Big)^{(n-2 \mu n)/(n+1)} G(\xi)
\end{align}
\end{subequations}
where we define the following normalizations:
\begin{equation}
\begin{split}
&\widetilde{x} \equiv \big(4 \pi R^2 h(0) \kappa_1 M^{-1}\big)^{-1/(n+1)} = R~[4 \pi h(0) C_1]^{-\frac{1}{n+1}} \\[0.75ex]
&\widetilde{v} \equiv f(0)\kappa_2 \widetilde{x}^{-\mu n} = \bigg(\frac{E}{M}\bigg)^{1/2} f(0)C_2[4 \pi h(0) C_1]^{\frac{\mu n}{n+1}} \\[0.75ex]
&\widetilde{\rho} \equiv h(0)\kappa_1  \widetilde{x}^{n} = \frac{M}{R^3}[(4\pi)^{-n} h(0) C_1]^{\frac{1}{n+1}}  \\[0.75ex]
&\widetilde{P} \equiv \frac{g(0)}{f(0)^2 h(0)} \widetilde{\rho} ~\widetilde{v}^2 = \frac{E}{R^3} g(0) [4 \pi h(0)]^{\frac{n(2\mu -1)}{n+1}}C_2^2C_1^{\frac{2\mu n+1}{n+1}}~,
\end{split}
\end{equation}
and $F(\xi),H(\xi), G(\xi) \text{ and } r(\xi)$ are numerical functions of the self-similar variable $\xi$, which is defined as:
\begin{equation}
\label{e:xi}
\xi = \bigg[\frac{m (n+1)}{4 \pi R^2 h(0) \kappa_1}\bigg]^{(-\mu n -1)/(n+1)} \kappa_2  ~t ~.
\end{equation}
The parameters $h(0), g(0) \text{ and } f(0)$ are constants that depend on $n$ and are found by solving the equations of the previous stage of shock passage numerically (see S60). Their values for $n=3$ and $n=3/2$ are given in Table \ref{t:fgh_values}. 

\begin{table}
\caption{Values of f(0), h(0) and g(0) for typical values of $n$ and $\gamma=5/3$.}\label{t:fgh_values}
\begin{center}
\begin{tabular}{ c c c c }
\hline
n & f(0) & h(0) & g(0) \\
\hline
3/2 & 0.72 & 26.98 & 1.26\\
3 & 0.64 & 53.29 & 1.57 \\
\end{tabular}
\end{center}
\end{table}

The functions $F(\xi),H(\xi),G(\xi)$ and $r(\xi)$ do not have an analytic solution on the whole range of $\xi$. However, they approach a single powerlaw profile at $\xi \gg 1$ (for a given $t$, this would refer to the outer layers of the envelope that have already expanded). Powerlaw fitting to the numerical solutions of $F(\xi), G(\xi), H(\xi) \text{ and } r(\xi)$ in the limit of $\xi\gg 1$ gives:
\begin{subequations}\label{e:FHGr_asymptotic}
\begin{align}
&F(\xi) = \alpha_F \\
&H(\xi) = \alpha_H \xi^{-1} \\
&G(\xi) = \alpha_G \xi^{-\gamma} \\
&r({\xi}) = \alpha_r \xi
\end{align}
\end{subequations}
where $\alpha_F, \alpha_H, \alpha_G \text{ and } \alpha_r$ are numerical parameters of the fit, whose values are given in Table \ref{t:alpha_values}. Using equation sets \eqref{e:planar_selfsim_hydro} and \eqref{e:FHGr_asymptotic} and the expression for $\xi$ in Eq \eqref{e:xi}, we write the expressions for $x, v,\rho$ and $P$ as functions of $m$ and $t$:
\begin{subequations}\label{e:conditions1}
\begin{align}
x(m,t)  & \!\begin{aligned}[t] &= \alpha_r \kappa_2 ~\widetilde{x}^{-\mu n} \Big(\frac{m}{M}\Big)^{-\frac{\mu n}{n+1}} t \end{aligned} \\[0.75ex]
\label{e:v_planar}
v(m) & \!\begin{aligned}[t] &= \alpha_F \widetilde{v}\Big(\frac{m}{M}\Big)^{-\mu n/(n+1)}  \end{aligned}\\[0.75ex]
\label{e:rho_planar}
\rho(m,t) & \!\begin{aligned}[t] &= \frac{\alpha_{H}}{\kappa_2} \widetilde{\rho} ~\widetilde{x}^{\mu n+1} \Big(\frac{m}{M}\Big)^{1+\frac{\mu n}{n+1}} t^{-1}
  \end{aligned}\\[0.75ex]
P(m,t)  & \!\begin{aligned}[t] &=\frac{\alpha_G}{\kappa_2^{\gamma}} \widetilde{P} ~\widetilde{x}^{\gamma(\mu n+1)} \Big(\frac{m}{M}\Big)^{\frac{n - 2\mu n + \gamma(\mu n+1)}{n+1}} t^{-\gamma}
\end{aligned}
\end{align}
\end{subequations}
Comparing Eq \eqref{e:v_planar} with Eq \eqref{e:v_xi}, we reproduce the known result by S60, showing that the velocity reaches an asymptotic value, which is roughly twice the initial velocity and is a function only of $m$. Furthermore, since the radius remains approximately constant during the planar phase and $x\propto t$, the density at a fixed mass coordinate evolves as $\rho\propto t^{-1}$. The evolution described in equation set \eqref{e:conditions1} does not start at $t=0$ for every $m$, since the width of a shell is effectively constant as long as its size has not doubled.
This earlier stage of the evolution is not described by the asymptotic solution presented in equation set \eqref{e:conditions1}.
%and starts at time $t_i(m) \sim d_0(m)/v_0(m)$ for a shell of mass $m$.
%We define the time a shell begins its planar evolution as the time when it has roughly doubled its intial width, and denote it as $t_i(m)$. We can therefore approximate 
%\begin{equation}
%t_i(m) \sim d_0(m)/v_0(m)
%\end{equation}
%where
%\begin{equation} \label{e:d_0}
%d_0 = \widetilde{x}\Big(\frac{m}{M}\Big)^{\frac{1}{n+1}}
%\end{equation}
%where $d_0(m)$ is the initial width of a shell of mass $m$ after shock passage and $v_0(m)$ is the initial velocity of that shell (right after shock passage, before the planar phase aceleration).

We use the solutions for $P$ and $\rho$ in equation set \eqref{e:conditions1} to find the evolution of the internal energy:
\begin{equation} \label{e:u_planar_adiabatic}
\begin{split}
u(m,t)& =  \frac{1}{\gamma -1}\frac{P}{\rho} =  \\[2ex]
&=  3 \frac{\alpha_G}{\alpha_H} ~\kappa_2^{-1/3} \frac{\widetilde{P}}{\widetilde{\rho}} ~\widetilde{x}^{\frac{1}{3}(\mu n+1)}\Big(\frac{m}{M}\Big)^{\frac{1-5\mu n}{3(n+1)}} t^{-1/3}.
\end{split}
\end{equation}
where we substituted $\gamma=4/3$ for a radiation dominated gas.
%This result can also be expressed as
%\begin{equation}
%\label{e:u_rho}
%u(m,t)= u_i(m)\Big[\frac{\rho(m,t)}{\rho_i}\Big]^{\gamma-1} = u_i(m)\Big(\frac{r^2d}{R^2d_i}\Big)^{-(\gamma-1)}
%\end{equation}
%with 
%\begin{equation}
%\label{e:u_i_explicit}
%u_i(m) = \frac{1}{\gamma -1}\frac{P(\xi_i)}{\rho(\xi_i)} \propto m^{\frac{-2\mu n}{n+1}} ~,
%\end{equation}
%where $d$ is the width of a shell of mass $m$ at time $t$, and $d_i$, $\rho_i$ and $u_i$ are the width of a shell, its density and the internal energy at $t=t_i(m)$ which corresponds to $\xi_i$. 
In the internal regions, diffusion is negligible and the internal energy changes only due to adiabatic cooling.

\begin{table}
\begin{center}
\begin{tabular}{c c c c c}
\hline
n & $\alpha_F$ & $\alpha_H$ & $\alpha_G$ & $\alpha_r$ \\
\hline
3/2 & 1.9 & 2.2 & 2.9 & 1.5  \\
3 &  1.7 & 1.5 & 1.8 & 1.2 \\
\end{tabular}
\end{center}
\caption{Fit parameters of $F(\xi)$, $H(\xi)$, $G(\xi)$ and $r(\xi)$ for typical values of $n$ and $\gamma=4/3$}
\label{t:alpha_values}
\end{table}

\section{Diffusion during the planar phase} \label{s:diffusion}
For the purpose of describing the diffusion of radiation inside the SN ejecta, it is useful to treat the envelope as a series of successive shells, where inside each shell the hydrodynamic properties, such as the density, velocity and radius do not vary a lot. 
Each shell is assigned a mass $m$, corresponding to that defined in Eq \eqref{e:m}, which is roughly the integrated mass measured from the edge of the stellar envelope down to the internal radius of the shell. According to this definition, the more massive the shell is, the more internal is its position inside the envelope. Each shell has a characteristic width $d = d_0+vt$, where $d_0$ is its pre-explosion width. The planar self-similar dynamics for each shell starts once its width has doubled, such that $d\sim x \sim vt \gg d_0$

In the previous section we found an expression for the internal energy when photon diffusion is negligible. In this section we find the energy profile that forms due to diffusion of photons through the ejecta, and by that compute the observed bolometric luminosity. We begin with an order of magnitude derivation of the luminosity shell, which is the source of the escaping photons, and then solve the equation that describes the energy evolution inside the envelope, to find an accurate self-similar solution for the energy and the post breakout shock cooling emission.
\subsection{The breakout and luminosity shells}
In the early stages after shock passage, the density and temperatures inside the envelope are very high and the matter is opaque to radiation. Since the temperatures are higher than $\sim 1$ eV, the gas is fully ionized, and the dominant opacity source is Thomson scattering. The diffusion time through the envelope therefore depends only on the mass of the shell and its width $d$: 
\begin{equation} \label{e:tdiff}
t_\diff= 3 \kappa_T \frac{m}{4\pi R^2} \frac{d}{c} ~,
\end{equation}
where $\kappa_T$ is Thomson opacity and $m$ is the mass of the shell.
%A shell of mass $m$ will have a roughly constant density for $t<t_i(m)$, which implies that the diffusion time from it is constant for $t<t_i(m)$, and increases linearly with $t$ for $t>t_i(m)$.
A photon originating in a shell of mass $m$ will be able to escape the envelope when the diffusion time from that shell is equal to its dynamical time, namely
\begin{equation} \label{e:luminosity_shell_condition}
t_{dyn} = t_\diff ~,
\end{equation}
where $t_\dyn = t$. Eq \eqref{e:luminosity_shell_condition} is equivalent to the condition $\tau=c/v$, where $\tau$ is the optical depth and $c$ is the speed of light. The shell that satisfies this condition is called the `breakout shell', and the time it takes photons to diffuse out of the bottom of that shell before expansion is the breakout time, $t_\bo$. The properties of the breakout shell are denoted by the subscript $_{\bo}$.
%Since $t_\bo$ is effectively the initial dynamical time of the breakout shell, then $t_\bo = t_\i(m_\bo)$.

Using Eq \eqref{e:tdiff} and the solutions for $\rho$ and $x$ in equation set \eqref{e:conditions1} (and setting $d\sim x$), we find the expression for the mass of the breakout shell:
\begin{equation}
\label{e:m_lum_naive}
m_{bo} = M \bigg(\frac{3 \kappa M}{4\pi R^2 c} \alpha_r \kappa_2 \widetilde{x}^{-\mu n}\bigg)^{\frac{n+1}{\mu n-n-1}}
\end{equation}
The breakout time is evaluated as the initial dynamical time of the breakout shell, namely $t_\bo = d_{0,\bo}/v_\bo$, where 
\begin{equation}\label{e:d_0}
d_0(m) = \widetilde{x}\Big(\frac{m}{M}\Big)^{\frac{1}{n+1}}
\end{equation}
is the initial width of a shell of mass $m$ before expansion and $v_\bo$ is the velocity of the breakout shell before the planar acceleration. Using equations \eqref{e:v_planar}, \eqref{e:m_lum_naive} and \eqref{e:d_0} we find the expression for the breakout time:
\begin{equation}
\label{e:ti_bo}
\begin{split}
t_\bo = \widetilde{x}^{1-\frac{\mu n(1+\mu n)}{\mu n-n-1}}\bigg(\frac{3 \kappa M}{4\pi R^2 c} \kappa_2 \alpha_r\bigg)^{\frac{1+\mu n}{\mu n-n-1}}
\end{split}
\end{equation}
The properties of the breakout shell at $t_\bo$ can be found by substituting the mass of the breakout shell in Eq \eqref{e:m_lum_naive} into equation set \eqref{e:conditions1}. The typical breakout properties for Red Supergiant (RSG), Blue Supergiant (BSG) and Wolf-Rayet (WR) explosions are given in Appendix \ref{ap:bo_properties}.
%As stated above, a certain shell will start its planar evolution after it has doubled its width. Before that stage, the specific energy of the shell is approximately constant. The dependence of $t_i$ on $m$ can be found using the velocity given in Eq \eqref{e:conditions1} and the initial shell width in Eq \eqref{e:planar_selfsim_hydro}:
%\begin{equation}
%\label{e:ti}
%t_i(m) \sim \frac{d_0(m)}{v(m)} = t_\bo\Big(\frac{m}{m_{bo}}\Big)^{(1+\mu n)/(n+1)} ~.
%\end{equation}
%Reverting the expression for $t_i(m)$, we can estimate the evolution of the shell that satisfies $m(t=t_i)$: 
%
%\begin{equation}
%\label{e:m_ti}
%m(t_i)  =  m_{bo} \Big(\frac{t}{t_\bo}\Big)^{(n+1)/(1+\mu n)}
%\end{equation}
%Since the luminosity shell remains roughly constant near $m_{bo}$, the shell $m(t_i)$ will always be deeper than $m_{lum}$ (or equal to it at $t_\bo$) and therefore $m_\lum$ will evolve according to equation set  \eqref{e:conditions1} for all $t>t_\bo$. 

At later times, the shell out of which the observed photons originate is called the "luminosity shell" and its properties are denoted by the subscript $_\lum$. The luminosity of a SN at time $t$ is dominated by the energy stored in the luminosity shell. External to the luminosity shell, at $m<m_\lum$, the luminosity is constant.
Equations \eqref{e:tdiff} and \eqref{e:luminosity_shell_condition} imply that the optical depth of each shell is constant in time during the planar phase. Therefore, there is only one shell that satisfies $\tau=c/v$ and naively, the breakout shell should serve as the luminosity shell throughout the entire planar phase. Although this is the picture that was accepted until now \citep[][hereafter NS10]{ns10}, it can be contradicted with simple arguments.
Given that the duration of the planar phase is much longer than the dynamical time of the breakout shell at breakout, the time will double itself many times until the beginning of the spherical phase, such that at some point the breakout shell will have lost roughly all of its energy, and photons from innermore shells will be able to diffuse out of the envelope. The luminosity shell will thus recede, though slowly, and expose deeper regions of the outflow.
To estimate the correction to the mass of the luminosity shell, consider a shell of mass $m$ that is internal to $m_{\bo}$, such that $t_\dyn(m)<t_\diff(m)$. A fraction of the energy in that shell is able to escape, and the energy that remains in it at time $t$ can be approximated as:
\begin{equation}\label{e:E_lum_approx}
E(t) \approx E_0 \Bigg(1 - \frac{t_\dyn(m)}{t_\diff(m)}\Bigg)^{\log\big(\frac{t}{d_0(m)/v(m)}\big)} ~,
\end{equation}
where $E_0$ is the initial internal energy in the shell and $\log \big(\frac{t}{d_0(m)/v(m)}\big)$ counts the number of dynamical times that have passed for that shell, since the beginning of expansion. The shell out of which photons effectively diffuse out, namely the luminosity shell, can be defined as the shell that satisfies $E(t)/E_0\sim 0.5$. Solving Eq \eqref{e:E_lum_approx} for this condition with $t_\dyn/t_\diff \ll 1$ we find an approximate expression for the mass of the luminosity shell:
\begin{equation} \label{e:m_lum_planar_approx}
m_\lum \approx m_\bo\bigg[2\cdot \log\bigg(\frac{t}{d_0(m)/v(m)}\bigg)\bigg]^{(n+1)/(n+1-\mu n)} ~.
\end{equation}
The luminosity shell satisfies $t<t_\diff$, but is nevertheless the shell out of which photons diffuse out to the observer.

In the next section, we derive a more rigorous and accurate expression for $m_\lum$ that agrees with the scaling in Eq \eqref{e:m_lum_planar_approx}.

\subsection{Self-Similar Solution for the radiation}
\label{s:energy_selfsim}
The equation that describes the evolution of the specific energy is:
\begin{equation}
\label{e:u_equation}
\frac{\partial u}{\partial t} = -\frac{\partial}{\partial m}(4\pi r^2 \mathcal{F}) - \frac{u}{3t}~,
\end{equation}
where
\begin{equation}
\label{e:F}
\mathcal{F} = \mathcal{D}\frac{\partial (u\cdot \rho)}{\partial x}
\end{equation}
is the flux and $\mathcal{D} = c/3\kappa \rho$ is the diffusion coefficient.
Eq \eqref{e:u_equation} can be solved with a self-similar solution. We want to transform $u$ and $m$ into new variables, such that the dependence on $t$ is eliminated from the equation and it becomes an ordinary differential equation (ODE). In this problem the luminosity shell serves as the self-similar coordinate, since it is the location from which photons effectively diffuse out, and the specific energy profile changes from an inner adiabatic profile to an outer diffusive profile. Eq \eqref{e:m_lum_planar_approx} is an estimate for the mass of the luminosity shell. A more formal definition is obtained by determining the correct transformation of the variable $m$. We represent the luminosity shell as an unknown function of the time, $m_\lum=F(t)$, and define new variables by normalizing $m$ and $u$ to the mass and the adiabatic energy of the luminosity shell, respectively:
%(namely $u(m_\lum,t)$, given by Eq \eqref{e:u_planar_adiabatic})
\begin{equation} \label{e:planar_normalization}
\begin{split}
&\widetilde{m} \equiv \frac{m}{F(t)} \\
&\widetilde{u} \equiv \frac{u}{u_{bo}\Big[F(t)/m_{bo}\Big]^{\frac{1-5\mu n}{3(n+1)}}\Big(\frac{t}{t_{bo}}\Big)^{-1/3}} ~.
\end{split}
\end{equation}
Plugging the new variables into Eq  \eqref{e:u_equation}, it can be shown that in order to transform Eq \eqref{e:u_equation} into an ODE, $F(t)$ needs to satisfy:
\begin{equation}
F'(t)\cdot \Big[F(t)\Big]^{-\mu n/(n+1)} \propto t^{-1}
\end{equation}
with the initial condition $F(t = t_\bo) = m_\bo$. The two conditions above give the solution for $F(t)$:
\begin{equation} \label{eq:Ft_planar}
m_\lum(t) = F(t) = m_{bo}\big[1+\log(t/t_\bo)\big]^{(n+1)/(n+1-\mu n)} ~,
\end{equation}
We note that although radiation leaks out of deeper shells, only the breakout shell satisfies $t = t_\diff$.

In the problem we are solving, the pre-diffusion profiles are represented by power laws as evaluated in equation set \eqref{e:conditions1}, which imply unphysical infinite values at $x=0$. In this representation, according to Eq \eqref{eq:Ft_planar} the first shell out of which energy leaks is located at $m=0$, which is satisfied when $t = t_\bo/e$. Therefore, between the time the first photons started leaking out of the envelope and $t=t_\bo$, the time has only roughly doubled itself. The considered power law profiles are obviously not realistic close to $x=0$, and serve as a good description of the real SN envelope only at small values of $m$ or large $t$ (i.e. $\xi\gg 1$).

\begin{figure}
 \centering
\includegraphics[width=0.9\columnwidth]{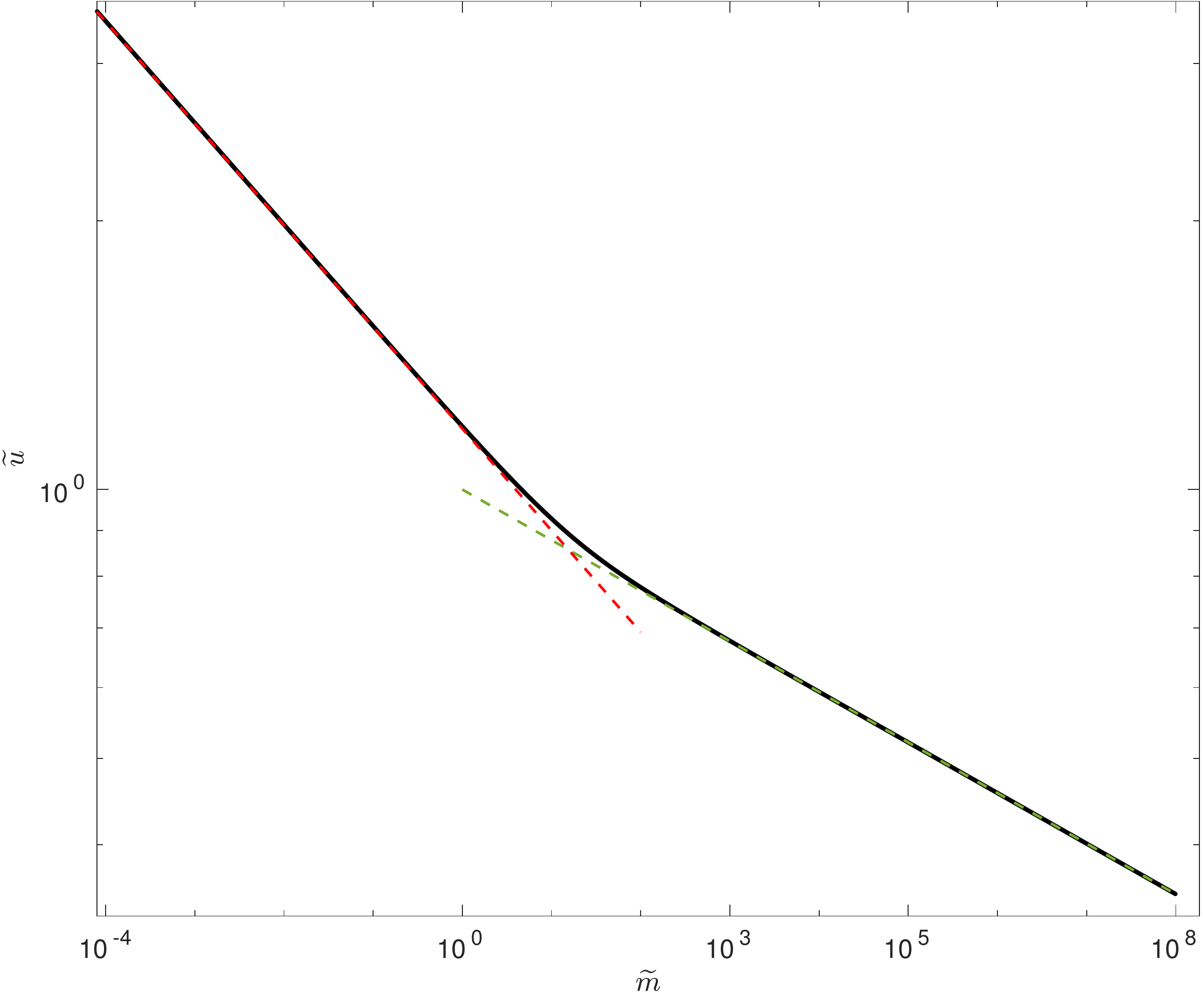} 
\caption{The self-similar solution of Eq \eqref{e:self_similar_planar} for the case of $n=3/2$. The black curve is the numerical solution, while the dashed lines are the analytic solutions for $\widetilde{m}\ll 1$ (red) and $\widetilde{m}\gg 1$ (green), according to Eq \eqref{e:u_tilde_planar}.}\label{f:u_planar}
\end{figure}

With the corrected expression for $m_\lum$, we perform the following variable transformation:
\begin{subequations} \label{e:self_similar_normalization}
\begin{align}
&\widetilde{m} \equiv \frac{m}{m_{bo}\Big[1+\log(t/t_\bo)\Big]^{\frac{n+1}{n+1-\mu n}}} \\
&\widetilde{u} \equiv \frac{u}{u_{bo}\Big[1+\log(t/t_\bo)\Big]^{\frac{1-5\mu n}{3(n+1-\mu n)}}\Big(\frac{t}{t_\bo}\Big)^{-1/3}} ~,
\end{align}
\end{subequations}
and obtain the self-similar form of Eq \eqref{e:u_equation}:
\begin{equation}\label{e:self_similar_planar}
\begin{split}
&\frac{\dd^2 \widetilde{u}}{\dd \widetilde{m}^2}\widetilde{m}^{1+\mu n/(n+1)}\frac{n+1}{\mu n} + \\[1.5ex]
& \frac{\dd  \widetilde{u}}{\dd \widetilde{m}}\bigg[2 \frac{n+1}{\mu n} \Big(1+\frac{\mu n}{n+1}\Big)\widetilde{m}^{\mu n/(n+1)} + \widetilde{m}\frac{n+1}{n+1-\mu n}\bigg] + \\[1.5ex]
 &\widetilde{u}\bigg[\Big(1 +\frac{\mu n}{n+1}\Big) \widetilde{m}^{\mu n/(n+1)-1}-\frac{1-5\mu n}{3(n+1-\mu n)}\bigg]= 0 ~.
 \end{split}
\end{equation}
This equation can be solved analytically in the limits of $\widetilde{m}\gg1$ and $\widetilde{m}\ll1$ by substituting a general power law solution of the form $\widetilde{u}=A \widetilde{m}^{\delta}$. We find:
\begin{equation} \label{e:u_tilde_planar}
\widetilde{u} =
\begin{cases}
\widetilde{m}^{(1-5 \mu n)/3(n+1)} &, \widetilde{m}\gg1\\
\\
A \widetilde{m}^{-\mu n/(n+1)}  &, \widetilde{m}\ll1
\end{cases}
\end{equation}
where $A$ is a numerical constant that is found by solving Eq \eqref{e:self_similar_planar} numerically. The solution for $u$ in the limit $m \gg m_{lum}$ is known accurately from Eq \eqref{e:u_planar_adiabatic} since it is purely adiabatic, and we use it as a boundary condition at $\widetilde{m}\gg 1$. Fitting the numerical solution for $A\widetilde{m}^{\delta}$ in the limits $\widetilde{m}\gg 1$ and $\widetilde{m}\ll 1$, we find that $A=1.17$ for $n=3/2$, and $A = 0.97$ for $n=3$. In Figure \ref{f:u_planar} we plot the numerical solution for $n=3/2$, and the analytic approximations at $\widetilde{m}\ll1$ and $\widetilde{m}\gg1$. 

Using equations \eqref{e:self_similar_normalization} and \eqref{e:u_tilde_planar} we write the solution for $u$ in the external parts of the ejecta:
\begin{equation}\label{e:u_planar_external}
\begin{split}
u(m\ll m_\lum) = &A u_\bo\bigg(\frac{m}{m_\bo}\bigg)^{-\frac{\mu n}{n+1}} \Big[1+\log(t/t_\bo)\Big]^{\frac{1-2\mu n}{3(n+1-\mu n)}}\\[2ex]
&\times\bigg(\frac{t}{t_\bo}\bigg)^{-1/3} ~.
\end{split}
\end{equation}
The solution at $\widetilde{m}\ll1$ is then used to compute the bolometric luminosity of the SN.

\section{The bolometric luminosity}\label{s:luminosity}
\subsection{Planar phase luminosity}
The bolometric luminosity during the planar phase is computed according to
\begin{equation}
L = 4\pi R^2\mathcal{F}~,
\end{equation}
where $\mathcal{F}$ was defined in Eq \eqref{e:F}. Using the self-similar solution for $u(m\ll m_\lum)$ in Eq \eqref{e:u_planar_external} together with the expression for $\rho$ in Eq \eqref{e:rho_planar} we can express the luminosity as:
\begin{equation} \label{e:L_planar}
\begin{split}
&L = A\frac{u_\bo m_\bo}{t_\bo} \bigg(\frac{n+1}{\mu n}\bigg)[1+ \log(t/t_\bo)]^{\frac{1-2\mu n}{3(n+1-\mu n)}}\\[2ex] &\times \bigg(\frac{t}{t_\bo}\bigg)^{-4/3} \propto
\begin{cases}
t^{-4/3} [1+ \log(t/t_\bo)]^{0.06} &, n=3/2 \\[2ex]
t^{-4/3} [1+ \log(t/t_\bo)]^{-0.01} &, n=3 ~.
\end{cases}
\end{split}
\end{equation}
%For typical properties of RSG and BSG, the luminosity of the planar phase is:
%\begin{equation}
%L(t) = 
%\begin{cases}
%2.2\cdot 10^{46}  \text{ erg}\cdot \text{s}^{-1} M_{15}^{-0.36} R_{500}^{2.46} E_{51}^{0.30} \kappa_{0.34}^{-1.06}\times \\[2ex] C_1^{-0.06} C_2^{0.60} [1+ log(t/t_\bo)]^{0.06} \Big(\frac{t}{t_\bo}\Big)^{-4/3}, n=3/2 \\[4ex]
%4.9\cdot 10^{45}  \text{ erg}\cdot \text{s}^{-1} M_{15}^{-0.33} R_{50}^{2.31} E_{51}^{0.34} \kappa_{0.34}^{-0.99}\times \\[2ex] C_1^{0.01} C_2^{0.68}~[1+log(t/t_\bo)]^{-0.01} \Big(\frac{t}{t_\bo}\Big)^{-4/3}, n=3  \\
%\end{cases}
%\end{equation}
The factor $[1+\log(t/t_\bo)]^{\frac{1-2\mu n}{3(n+1-\mu n)}}$ can thus either slow down or accelerate the decrease of the luminosity, depending on the value of $n$. However, the power of the logarithmic correction is very small, and does not have a noticeable effect on the observed luminosity of typical SN events.
The reason for the weak response of the luminosity to the significant increase in $m_\lum$ is the flat profile of the internal luminosity in the adiabatic part of the envelope (denoted $L_{\text{int}}(m)$, the energy per unit time that crosses a shell inside the SN envelope). Using the expression for the adiabatic evolution of $u$ in Eq \eqref{e:u_planar_adiabatic} we can write the profile of the internal luminosity in shells deeper than the luminosity shell:
\begin{equation}
L_{\text{int}}(m) \approx \frac{u(m_\lum<m)\cdot m}{t_\diff} \propto t^{-4/3} m^{\frac{1-2\mu n}{3(n+1)}} ~.
\end{equation}
For $n=3/2$ the power of $m$ is $\sim 0.06$ and for $n=3$ it is $-0.01$, implying that the spatial profile of the luminosity is rather flat. Therefore, the increase in $m_\lum$ can have only a scarce effect on the luminosity.

%The luminosity originates in regions where $t<t_\diff$, and the dependence of $t_\diff$ on $m$ in the denominator of the above expression causes the luminosity profile to be rather flat.

\subsection{Planar-Spherical Transition}\label{s:L_transition}
The planar phase ends when the ejecta radius has roughly doubled itself. The dynamics then transition into the spherical phase, where the SN radius is no longer constant, but evolves homologously as $r\sim v\times t$. Since the ejecta velocity is a function only of the mass (Eq \ref{e:v_planar}), lower mass shells enter the spherical phase earlier than more massive, internal shells.
The spherical time of each shell can be approximated as follows:
\begin{equation}
\label{e:t_s}
t_\text{spherical}(m) \approx \frac{R}{v(m)} = t_\s \bigg(\frac{m}{m_\bo}\bigg)^{\mu n/(n+1)} ~,
\end{equation}
where we define $t_\s \equiv R/v_\bo$ as the time the breakout shell enters the spherical phase.

At $t_\s$, the luminosity shell has reached a mass that is several times greater than $m_\bo$. We denote this mass coordinate as $m_\p$, and express it using Eq \eqref{eq:Ft_planar}:
\begin{equation}\label{e:m_pl}
m_\p \equiv m_\lum(t_\s) = m_\bo \Big[1+\log(t_\s/t_\bo)\Big]^{\frac{n+1}{n+1-\mu n}} \approx 10~ m_\bo~.
\end{equation}
Since a shell's radius starts increasing at $R/v(m)<t$, the optical depth of a shell of mass $m$ is no longer constant, and the shell that satisfies $\tau=c/v$ recedes inwards from the breakout shell, while initially it is shallower than $m_\p$. 
Naively, this means that photons diffuse out of shells that have already radiated away their energy during the planar phase. 
However, since the time has doubled itself many times since the begining of expansion, photons from inner shells were able to diffuse out and re-fill the shells external to $m_\p$, forming a diffusive energy profile. This is exactly the source of the logarithmic correction to $m_\lum$. 

At $t \lesssim t_\s$, shells internal to $m_\p$ evolve adiabatically, and no radiation diffuses from the inside, until $m_\lum=m_\p$. This is true since during the transition phase, the time has not doubled itself enough times to allow for radiation from internal regions to diffuse out, and the energy profile at $m_\pl<m$ is roughly constant.

\begin{figure}
 \centering
\includegraphics[width=1\columnwidth]{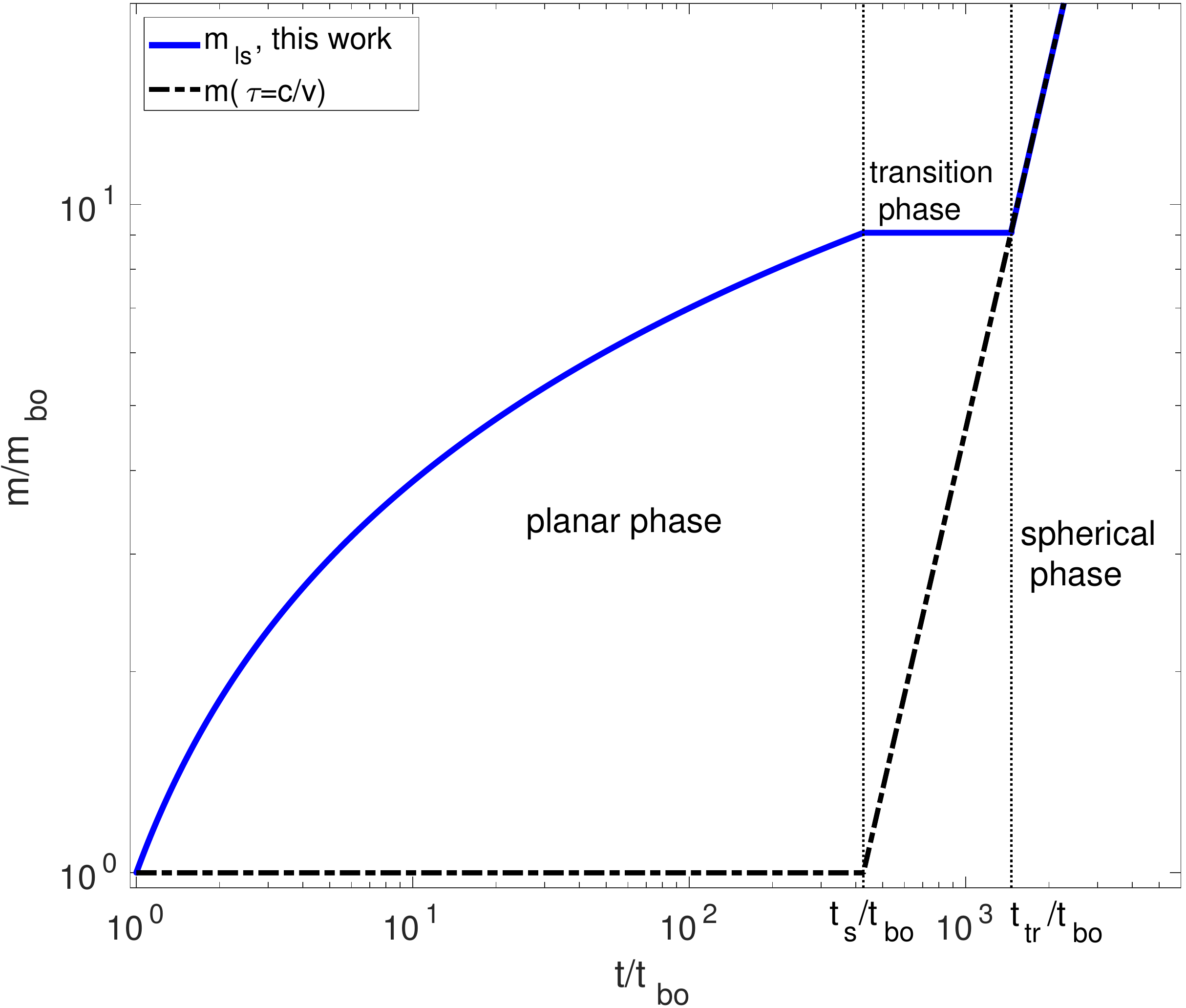} 
\caption{The luminosity shell including the logarithmic correction for $n=3/2$ (solid blue line) and the naive luminosity shell, which satisfies $\tau=c/v$ throughout the evolution (black dash-dotted line). The luminosity shell grows logarithmically with time until it enters the transition phase at $t_\s=R/v_\bo$. It then remains constant until it satisfies $\tau=c/v$ and enters the fully spherical phase. At the end of the planar phase, $m_\lum\sim ~10 m_\bo$.} \label{f:m_planar}
\end{figure}

The shell that was the luminosity shell at $t_\s$ remains so until it satisfies $\tau=c/v$, since it is the source of the observed luminosity. Afterwards, the luminosity shell is located where $\tau=c/v$.
%The shell that generated the photons that are the source of the observed luminosity at $t_\s \lesssim t$ is the shell that was the luminosity shell as $t_\s$, and therefore it remains the luminosity shell until it satisfies $\tau=c/v$.
%Afterwards, the luminosity shell is located at the location $\tau=c/v$ probes shells that are the energy source of the observed radiation as they generate their own photons (and do not fill by diffusion of photons from inner shells), and the shell in which $\tau=c/v$ is the luminoisty shell.
As a result, before the radiation enters the `nominal' spherical phase, it first goes through a transition phase, during which the luminosity is governed by the energy that diffused out of the luminosity shell at $t_\s$.
In Figure \ref{f:m_planar} we compare the luminosity shell defined in this work to the previously perceived `naive' luminosity shell, which is located at $m(\tau=c/v)$ throughout the evolution.
%\begin{equation}
%\label{e:t_bo_spherical}
%t_{\s,\bo} \approx \frac{R}{\alpha_F \widetilde{v}}\Big(\frac{3 \kappa}{c} \kappa_2 \alpha_H \alpha_r^2 \widetilde{\rho}~ \widetilde{x}^{\mu n -1}\Big)^{-\frac{\mu n}{n+1-\mu n}} ~.
%\end{equation}

In order to find the bolometric luminosity during the transition phase, we first find the expressions for the relevant hydrodynamic properties during their spherical evolution. 

The radii of the shells within the spherical phase follow a homologous expansion, and have the following dependence on $m$ and $t$:
\begin{equation}
\label{e:r_spherical}
r(m,t) \approx v(m)\cdot t = v_\bo \cdot t \bigg(\frac{m}{m_\bo}\bigg)^{-\mu n/(n+1)} 
\end{equation}
The evolution of $r$ together with conservation of mass imply that the density of a shell drops as $\rho \propto t^{-3}$ starting at $R/v(m)$:
\begin{equation}
\label{e:rho_sp}
\begin{split}
\rho(m) &= \rho(m,t=R/v(m)) \bigg(\frac{t}{R/v(m)}\bigg)^{-3} \\[1ex]
&= \rho_\bo \bigg(\frac{m}{m_\bo}\bigg)^{1+3\mu n/(n+1)} \bigg(\frac{t_\s}{t_\bo}\bigg)^{-1} \bigg(\frac{t}{t_\s}\bigg)^{-3}~.
\end{split}
\end{equation}
With the above expressions for $\rho \text{ and } r$, we can estimate the mass of the shell that satisfies $\tau=c/v$ during the spherical phase by requiring that $t_{diff}=t_{dyn}$. The diffusion time during the spherical phase can be written as
\begin{equation}
t_\diff \approx 3\kappa_T \frac{m}{4\pi r c}
\end{equation}
and therefore
\begin{equation}\label{e:m_lum_sph}
m(\tau=c/v) = m_\bo \bigg(\frac{t}{t_\s}\bigg)^{\frac{2(n+1)}{n+1+\mu n}}~.
\end{equation} 
The specific energy density at $m_\p<m$ can be found using the solutions for $u$ and $\rho$ at $t=R/v(m)$ in equations \eqref{e:rho_planar} and \eqref{e:u_planar_adiabatic}, and the expression for $\rho$ in the spherical phase in Eq \eqref{e:rho_sp}:
\begin{equation}\label{e:u_adiabatic_spherical}
\begin{split}
&u(m_\p<m) = \\ 
&=u(m_\p<m,t=R/v(m))\bigg[\frac{\rho(m,t)}{\rho(m,t=R/v(m))}\bigg]^{\gamma-1}\\[2ex]
& = u_\bo(t_\s)\bigg(\frac{m}{m_\p}\bigg)^{\frac{1-3\mu n}{3(n+1)}}\bigg(\frac{t}{t_\s}\bigg)^{-1} \Big[1+\log(t_\s/t_\bo)\Big]^{\frac{1-3\mu n}{3(n+1-\mu n)}}
\end{split}
\end{equation}
The specific energy profile at $m<m_\p$ was formed by diffusion at the end of the planar phase. Using Eq \eqref{e:u_planar_external}, we find:
\begin{equation}\label{e:u_diffusive_spherical}
\begin{split}
&u(m<m_\p) \\[2ex]
&=u(m<m_\p,t=R/v(m))\bigg[\frac{\rho(m,t)}{\rho(m,t=R/v(m))}\bigg]^{\gamma-1}\\[2ex]
& = u_\bo(t_\s)\bigg(\frac{m}{m_\bo}\bigg)^{-\frac{\mu n}{3(n+1)}}\bigg(\frac{t}{t_\s}\bigg)^{-1} \Big[1+\log(t_\s/t_\bo)\Big]^{\frac{1-2\mu n}{3(n+1-\mu n)}}
\end{split}
\end{equation}
The luminosity in the transition phase can be approximated using equations \eqref{e:m_lum_sph} and \eqref{e:u_diffusive_spherical}:
\begin{equation}\label{e:L_transition}
\begin{split}
L(t) \approx &\frac{u(\tau=c/v) \cdot m(\tau=c/v)}{t} \\[2ex]
&= \frac{u_\bo m_\bo}{t_\s}\bigg(\frac{t_\s}{t_\bo}\bigg)^{-1/3}\bigg(\frac{t}{t_\s}\bigg)^{-\frac{8 \mu n}{3(n+1-\mu n)}} ~.
\end{split}
\end{equation}
%\begin{equation}\label{e:L_transition}
%\begin{split}
%L(t) \approx&\frac{E_\lum(R/v_\lum)}{t}\bigg[\frac{t}{R/v_\lum}\bigg]^{-1}  \\[2ex]
%&=~\frac{E_\bo(t_\bo) (t_\s/t_\bo)^{-1/3}}{t_\s}\Big(\frac{t}{t_\s}\Big)^{-\frac{8\mu n}{3(n+1+\mu n)}}
%\end{split}
%\end{equation}
%\begin{equation}\label{e:L_transition}
%\begin{split}
%L(t) \approx &\frac{E(m_\lum)}{t}\bigg[\frac{t}{t_\s(m_\lum)}\bigg]^{-1} \propto t^{-\frac{8\mu n}{3(n+1+\mu n)}} \propto \\[2ex]
%&\propto 
%\begin{cases}
%t^{-0.27} &, n=3/2 \\\\
%t^{-0.33} &, n=3 ~,
%\end{cases}
%\end{split}
%\end{equation}
At $\tau<c/v$, the specfic energy profile is formed by photon diffusion and is given by $u = L \tau/4\pi r^2 c \rho$, where  $\tau\approx 3\kappa_T m/4\pi r^2$. Using equations \eqref{e:rho_planar}, \eqref{e:r_spherical} and \eqref{e:L_transition} we have:
\begin{equation}\label{e:u_external_transition}
u(\tau<c/v) = u_\bo(t_\s) \bigg(\frac{m}{m_\bo}\bigg)^{1+\frac{4\mu n}{n+1}} \bigg(\frac{t}{t_\s}\bigg)^{-4-\frac{8\mu n}{3(\mu n+n+1)}} ~.
\end{equation}
The transition to the spherical phase starts at $t_\s$ and ends when $m(\tau=c/v) = m_\p$. This time is denoted as $t_\tr$, and is estimated using equations \eqref{e:m_pl} and \eqref{e:m_lum_sph}:
\begin{equation} \label{e:t_transition}
t_\tr =t_\s~[1+\log(t_\s/t_\bo)]^{\frac{n+1+\mu n}{2(n+1-\mu n)}} \approx 3~t_\s ~.
\end{equation}
Past $t_\tr$, the dynamics of the radiation are fully spherical, and have been treated in several previous works \citep[NS10,][]{rabinak11}.

\section{The observed temperature} \label{s:T_obs}
As implied from Eq \eqref{e:L_planar}, the luminosity is only mildly affected by the logarithmic correction to $m_\lum$ during the planar phase. Nevertheless, the increase in $m_\lum$ means that the radiation originates in regions of higher optical depth and higher density, where the ability of the radiation to thermalize with the gas is significantly better than in the breakout shell. This will have a non negligible effect on the observed temperature.

Throughout this chapter, we use the method and terminology of NS10 to estimate the observed tempeature of the SN. Instead of assuming thermal equilibrium, they take into account the thermal coupling between the radiation and gas. The photon-electron coupling is achieved via several physical processes: free-free, bound-bound, bound-free emission and absorption, and also Compton and inverse Compton scattering. At the high temperature conditions that prevail during the planar phase, most of the hydrogen atoms are ionized ($T\gg 1 \text{ eV}$), such that bound-bound and bound-free absorption opacities are negligible compared to that of free-free. We thus take free-free absorption and emission as the dominant photon production processes, which can be complemented by Compton upscattering of lower energy photons.
During the thermalization process, radiation with an initial temperature of $T>T_\BB$ (where $T_\BB$ is its blackbody temperature) cools by sharing its energy with other photons, via emission and absorption from electrons. In order for a shell to achieve thermal equilibrium, it needs to produce the following number of photons of energy $3 k T_\BB$:
\begin{equation}
n_\BB \approx \frac{aT_\BB^4}{3kT_\BB} ~.
\end{equation}
where $k$ is Boltzmann's constant, $a$ is the radiation constant, and the blackbody temperature can be expressed as
\begin{equation}\label{e:T_BB}
T_\BB = \Big(\frac{u \cdot \rho}{a}\Big)^{1/4} ~.
\end{equation}
In order to determine whether a certain shell is capable of achieving thermal equilibrium, we use the following definition for the thermal coupling coefficient:
\begin{equation} \label{e:eta_def}
\eta \equiv \frac{n_{\text{BB}}}{\min\{t,t_{\text{diff}} \} \dot{n}(T_\text{BB})}
\end{equation}
where $\dot{n}(T_\text{BB})$ is the production rate of photons of energy $3k T_\BB$, and $\min\{t,t_{\text{diff}} \}$ is the time a photon spends in the shell, and is equal to $t_\diff$ for shells satisfying $m<m_\bo$ and to $t$ for shells with $m>m_\bo$ during the planar phase.
The dominant process for photon production is considered to be free-free emission, for which $\dot{n}(T_\text{BB})$ in Eq \eqref{e:eta_def} is the free-free photon production rate:
\begin{equation} \label{e:n_dot}
\dot{n}_{\text{ph,ff}} = 3.5 \times 10^{36}\text{s}^{-1} \text{cm}^{-3} \rho ^2 T_\BB^{-1/2} ~,
\end{equation}
for which Eq \eqref{e:eta_def} becomes
\begin{equation} \label{e:eta_ff}
\eta_{\text{ff}} = \frac{7 \times 10^5 \text{s}}{\min\{t,t_{\text{diff}} \}} \bigg(\frac{\rho}{10^{-10}  \text{g cm}^{-3}}\bigg)^{-2} \bigg(\frac{k T_\BB}{100 \text{eV}}\bigg)^{7/2} ~.
\end{equation}
If photon production is the only process that thermalizes the radiation, $\eta<1$ is the condition for thermal equilibrium in a shell.
However, the definition of $\eta$ in Eq \eqref{e:eta_def} does not take into account the effect of inverse Compton scattering.
Photons that were created with an energy $h\nu<3kT$ can contribute to thermalization if they are Compton upscattered to $h\nu\sim 3kT$, where $h$ is Planck's constant.
The lowest energy photon that can contribute to the spectrum at $3kT$ is limited by two main processes: free-free self-absorption and photon diffusion out of the shell. Free-free self-absorption causes the number of available photons for Comptonization to decrease by a factor of $\nu^2$. The self-absorption frequency, $\nu_\sa$ is found by equating the RJ energy density to the energy density produced by free-free emission (which depends on the time a photon spends in a shell), and is equal to
\begin{equation}
\begin{split}
h\nu_\sa \approx &10~ \text{eV} ~g_\ff^{1/2} \bigg(\frac{n}{10^{15}~\text{cm}^{-3}}\bigg)^2~ \bigg(\frac{T}{100~\text{eV}}\bigg)^{-3/4}\\[2ex]
&\times \bigg(\frac{\min\{t,t_\diff\}}{1~\text{s}}\bigg)^{1/2} ~,
\end{split}
\end{equation}
where we used the fact that at the RJ tail $h\nu \ll kT$.
On the other hand, another limit comes from the fact that scattered photons must interact with the electrons a sufficient amount of times to reach an energy of $3kT$ before escaping the shell by diffusion. The minimal frequency that satisfies this requirement is
\begin{equation}
\begin{split}
h\nu_{\diff} = k T \exp \Bigg\{-\frac{k T}{m_e c^2}\frac{\tau c}{v}\Bigg\}~.
\end{split}
\end{equation}
The lowest energy that can be Comptonized, $h \nu_\min$, is then determined by 
\begin{equation} \label{e:nu_min}
h \nu_\min = \max\{h \nu_\diff,h \nu_\sa\} ~.
\end{equation}

The Comptonization parameter, $\xi(T)$, is the ratio of photons at energy $3kT$ created by Comptonization, relative to free-free processes:
\begin{equation} \label{e:xi_def}
\xi(T) \approx \text{max}\{1,1+0.5 \cdot  \log(y_{\text{max}})(1.5+\log(y_{\text{max}})\}
\end{equation}
where 
\begin{equation}
y_{\text{max}} \equiv \frac{k T}{h \nu_{\text{min}}} ~.
\end{equation}
%\begin{equation}
%y_{\text{max}} \equiv \frac{k T}{h \cdot\text{max}\{\nu_{\text{min}},\nu_{\text{sa}}\}} ~.
%\end{equation}
%(notice the slight difference from NS10 in the definition of $\xi$).
Note that $\xi$ is an implicit function of $T$.
Furthermore, the logarithmic dependence of $\xi$ on $y_{\text{max}}$ resolves the exponential sensitivity of $h \nu_\diff$.

A shell in which $\eta/\xi<1$ implies that the radiation has had enough time to thermalize with the gas during the available interaction time, and has achieved a temperature of $T_\BB$. However, when $\eta/\xi>1$, the shell did not manage to generate enough photons with an energy of $3k T_\BB$ and did not reach thermal equilibrium.
It can be shown that $\eta$ is an increasing function of $r$, and therefore if a certain shell is out of thermal equilibrium, so are all of the shells external to it. This implies that the outermost shell that is able to achieve thermal equilibrium satisfies $\eta/\xi = 1$. This shell is termed the `colour shell', and its location is more commonly known as the `thermalization depth'. External to the colour shell, the gas can no longer modify the energy of the radiaion, and from that point the observed temperature is constant and satisfies $T_\cl=T_\BB(\eta/\xi=1)$.

%In shells internal to the luminosity shell, the energy evolves adiabatically according to Eq \eqref{e:u_planar_explicit} and also $t<t_\diff$, such that the thermal coupling coefficient is:
%\begin{equation} \label{e:eta_planar_adiabatic}
%\eta(m_\lum<m,t) = \eta_\bo (t_\bo)\Big(\frac{m}{m_\bo}\Big)^{-\frac{20+27n+62\mu n}{24(n+1)}} \Big(\frac{t}{t_\bo}\Big)^{-1/6}
%\end{equation}
%\begin{equation} \label{e:eta_planar_adiabatic}
%\begin{split}
%\eta(m,t) =& \eta_\bo (t_\bo)\Big(\frac{m}{m_\bo}\Big)^{-\frac{20+27n+62\mu n}{24(n+1)}} \Big(\frac{t}{t_\bo}\Big)^{-1/6}  \\[2ex] \propto
%&\begin{cases}
%m^{-1.30}t^{-1/6}  \qquad , n=3/2 \\[2ex]
%m^{-1.42} t^{-1/6} \qquad , n=3 ~. \\
%\end{cases}
%\end{split}
%\end{equation}

If the luminosity shell is out of thermal equilibrium, it is the only shell that affects the observed temperature since the energy of the photons cannot be changed as they propagate outwards. The observed temperature will thus be the typical energy of the photons in the luminosity shell:
\begin{equation}\label{e:T_obs}
T_{\text{obs}} =
\begin{cases}
\frac{\eta_\lum^2}{\xi_\lum^2} ~ T_{\text{BB,\lum}} \qquad ,\eta_\lum/\xi_\lum>1\\
\\
T_\BB(\eta=1)\quad ,\eta_\lum/\xi_\lum<1 ~.
\end{cases}
\end{equation}
%\begin{equation} \label{e:eta_planar_diffusive}
%\begin{split}
%&\eta(m_\bo < m < m_\lum) = \eta_\bo(t_\bo)~\Big(\frac{t}{t_\bo} \Big)^{-1/6}  \Big(\frac{m}{ m_\bo} \Big)^{-\frac{16\mu n +9n +9}{8(n+1)}}  \times \\
%&[1+log(t/t_\bo)]^{\frac{7(2\mu n-1)}{24(\mu n - n -1)}} \propto \\[0.90ex] \propto
%&\begin{cases}
%~ t^{-1/6}~[1+log(t/t_\bo)]^{0.06}~ m^{-1.35}  , \quad n=3/2\\[0.75ex]
%\\
%~t^{-1/6}~[1+log(t/t_\bo)]^{-0.01}~ m^{1.41},\quad n=3
%\end{cases}
%\end{split}
%\end{equation}
%
%In shells lying exernal to the breakout shell, the time available for thermalization is no longer $t$ but $t_\diff$, and their thermal coupling coefficient is:
%\begin{equation}\label{e:eta_planar_external_to_mbo}
%\begin{split}
%&\eta(m < m_\bo) \\[2ex] 
%&\propto~t^{-1/6}[1+log(t/t_\bo)]^{\frac{7(2\mu n-1)}{24(\mu n - n -1)}} m^{\frac{\mu n -8n -8}{8(n+1)}}  \\[2ex] \propto
%&\begin{cases}
%\ ~t^{-1/6}~[1+log(t/t_\bo)]^{0.06}~ m^{-2.24}  , \quad n=3/2\\[0.75ex]
%\\
%~t^{-1/6}~[1+log(t/t_\bo)]^{-0.01}~ m^{-2.27},\quad n=3
%\end{cases}
%\end{split}
%\end{equation}
The observed emission from a SN can be in or out of thermal equilibrium, depending on the progenitor properties. High energy explosions in compact stars tend to be out of thermal equilibrium. In the following subsections we find the evolution of $T_\obs$ as a function of time in the planar and transition phases, for both cases.
\subsection{$T_\obs$ during the planar phase} \label{s:T_obs_planar}
The observed temperature during the planar phase is determined by the ability of the radiation to thermalize at $m\leq m_\lum$. We assume here that thermalization occurs at $m_\bo< m \leq m_\lum$, such that the available time for thermalization in a shell is $t$. The dependence of $\eta$ on $m$ and $t$ is found using equations \eqref{e:rho_planar}, \eqref{e:u_planar_external}, \eqref{e:T_BB} and \eqref{e:eta_ff}:
\begin{equation} \label{e:eta_planar_diffusive}
\begin{split}
&\eta(m_\bo < m < m_\lum) = \eta_\bo(t_\bo)~\bigg(\frac{t}{t_\bo} \bigg)^{-1/6} \\[2ex]
& \times \bigg(\frac{m}{ m_\bo} \bigg)^{-\frac{16\mu n +9n +9}{8(n+1)}} [1+\log(t/t_\bo)]^{\frac{7(2\mu n-1)}{24(\mu n - n -1)}}
\end{split}
\end{equation}
Whether the observed radiation is thermalized depends on the values of $\eta$ and $\xi$ in the luminosity shell.
The expression for $\eta_\lum$ is found by substituting the solution for $m_\lum$ from Eq \eqref{eq:Ft_planar} into Eq \eqref{e:eta_planar_diffusive}:
\begin{equation} \label{e:eta_planar_lum}
\begin{split}
\eta_\lum(t) = &~ \eta(t_\bo) \bigg(\frac{t}{t_\bo}\bigg)^{-1/6} [1+\log(t/t_\bo)]^{-\frac{20+27n+62\mu n}{24(n+1-\mu n)}} \\[2ex] \propto
&\begin{cases}
t^{-1/6} [1+\log(t/t_\bo)]^{-1.47} \quad \qquad , n=3/2 \\[2ex]
t^{-1/6} ~[1+\log(t/t_\bo)]^{-1.66} \quad \qquad , n=3 ~.
\end{cases}
\end{split}
\end{equation}
The thermal coupling of the luminosity shell thus significantly improves owing to the logarithmic increase of $m_\lum$. The luminosity shell probes regions of higher density (by the end of the planar phase, the density of the luminosity shell is $\sim 10$ times higher than in the breakout shell), where the photon production rate increases accordingly as $\dot{n}\propto \rho^2$ (Eq \ref{e:n_dot}).

Since $\xi$ is an implicit function of $T$, the value of $\xi_\lum$ is ought to be computed numerically.
Using equations  \eqref{e:xi_def} and \eqref{e:eta_planar_lum} we can determine whether the observed radition is in thermal equilibrium or not, which will define the nature of the solution according to Eq \eqref{e:T_obs}.
%We note that if the ratio $\eta_\lum/\xi_\lum<1$, the radiation is in thermal equilibrium even when $\eta_\lum>1$.
\subsubsection{$T_\obs$ during the planar phase for $\eta_\lum/\xi_\lum >1$} \label{s:T_planar_noneq}
%For a definition of $\xi(T)$, see Eq(12) in \cite{ns10}.
%The above scaling indicates that during the planar phase, the thermal coupling increases with time and mass. The value of $\eta$ for $m=m_\bo$ (i.e., the normalization of Eq \eqref{e:eta_planar}) is different to the one found in \cite{ns10}. In this work, we treat the breakout shell as if it had been compressed by the shock, and ignore the dynamics of the shock breakout itself. This representation is obviously not accurate. However, our treatment is correct for all shells with $m>m_\bo$, and according to the logarithmic correction for the luminousity shell (Eq \eqref{eq:Ft_planar}), the shells that dominate the emission during the planar phase are internal to the breakout shell. Therefore, the above expression for $\eta$ is correct for the luminosity shell as it increases beyond the breakout shell. Therefore, Eq \eqref{eq:eta_planar} implies that shells with mass $m\sim m_\bo$ are not expected to be in thermal equilibrium. Nevertheless, the strong dependence of $\eta$ on $m$ can significantly decrease the value of $\eta$ as the luminosity shell recedes slowly into deeper layers. 
%If we substitude the expression for $m_\lum$ into Eq \eqref{eq:eta_planar}, we find that $\eta_\lum$ strongly depends on the logarithmic correction:
If $\eta_\lum/\xi_\lum>1$, the luminosity shell, and all shells external to it are out of thermal equilibrium. According to equations \eqref{e:T_obs} and \eqref{e:eta_planar_lum}, when the luminosity shell is out of thermal equilibrium $T_\obs$ behaves as
\begin{equation} \label{e:T_obs_planar}
\begin{split}
&T_{\text{obs}}(\eta_\lum/\xi_\lum>1) = \\[2ex]&  T_{\BB,\bo}(t_\bo)\frac{\xi_\lum(T)^2}{\xi_\lum(T_\bo)^2} \bigg(\frac{t}{t_\bo}\bigg)^{-2/3} [1+log(t/t_\bo)]^{\frac{16\mu n +6n+4}{3(\mu n-n-1)}} \\[2ex] \propto
&\begin{cases}
\xi_\lum(T)^{-2} ~t^{-2/3}[1+\log(t/t_\bo)]^{-2.64}  , \quad n=3/2\\
\\
\xi_\lum(T)^{-2} ~t^{-2/3}[1+\log(t/t_\bo)]^{-3.02},\quad n=3 ~,
\end{cases}
\end{split}
\end{equation}
where $T_\bo$ is the observed temerature at breakout.
The solution of $T_\obs$ is obtained by solving the implicit Eq \eqref{e:T_obs_planar} using the expression for $\xi(T)$ in Eq \eqref{e:xi_def}.
%Since $\xi(T)$ is itself a function of the temperature, $T_\obs$ is not a pure power law, and its dependence on time needs to be computed numerically.
Contrary to $L_\bol$, the dependence of $T_\obs$ on the logarithmic factor is not negligible, and the radiation approaches thermal equilibrium faster than previously thought. Equilibrium is achieved once $T_\obs=T_{\BB,\lum}$.

The expression for $T_\obs$ derived in Eq \eqref{e:T_obs_planar} is valid only while $T<50 \text{KeV}$. Above that temperature, relativistic effects such as pair production start being important \citep[]{weaver76,svensson84}. In that case, pair production will lead to a lower temperature than what Eq \eqref{e:T_obs} predicts. Therefore, if Eq \eqref{e:T_obs_planar} results in temperatures higher than $50 \text{KeV}$, it is not correct and the temperature should be computed using relativistic shock breakout models \citep{nakar12,budnik10}.
%Equation (13) neglects relativistic effects (pair production,
%relativistic bremsstrahlung, etc.) and is, therefore, accurate only
%for T  50 keV (Weaver 1976; Svensson 1984). We do
%not calculate here the exact temperature when Equation (13)
%predicts higher temperature. In this case, pair production will
%result in a lower temperature than Equation (13) predicts. In
%SNe, where the velocities are at most mildly relativistic, the
%temperature should fall within the range of soft gamma-ray
%detectors, ∼50–200 keV, when pair production is important
%(Katz et al. 2010).
\subsubsection{$T_\obs$ during the planar phase for $\eta_\lum/\xi_\lum<1$} \label{s:T_planar_eq}
When the luminosity shell is in thermal equilibrium, i.e. $\eta_\lum/\xi_\lum<1$, the observed temperature is determined in the colour shell, which satisfies $\eta=1$ (throughout this paper we assume that the colour temperature is low enough such that Compton scattering is not impotant in the colour shell, and $\xi_\cl\sim 1$). Assuming that the colour shell is external to the luminosity shell but internal to the breakout shell, $t<t_\diff$ in the colour shell and $t$ is the available time for the radiation to thermalize. 
Using Eq \eqref{e:eta_planar_diffusive}, we find the mass of the colour shell by solving $\eta=1$:
\begin{equation}\label{e:mcl_planar}
\begin{split}
m_\cl = & m_\bo \Big[\eta_\bo(t_\bo)\Big]^{\frac{8(n+1)}{16\mu n+9n+9}} \bigg(\frac{t}{t_\bo}\bigg)^{\frac{4(n+1)}{3(16\mu n+9n+9)}} \\[2ex]
&\times \Big[1+\log(t/t_\bo)\Big]^{\alpha} \\[2ex]
&\propto\begin{cases}
t^{-0.12} [1+\log(t/t_\bo)]^{0.04}  , \quad n=3/2\\
\\
t^{-0.12} [1+\log(t/t_\bo)]^{-0.01}, \quad n=3 
\end{cases}
\end{split}
\end{equation}
where
\begin{equation}
\alpha = \frac{7(n+1)(2\mu n-1)}{3(\mu n-n-1)(16\mu n +9n+9)} ~.
\end{equation}
The colour temperature is the blackbody temperature of the colour shell.
We thus use equations  \eqref{e:rho_planar} , \eqref{e:u_planar_external}, \eqref{e:T_BB} and \eqref{e:mcl_planar} to find $T_\cl$:
\begin{equation}\label{e:T_cl_planar}
\begin{split}
T_\cl = &T_\BB(m_\bo,t_\bo) \Big(\frac{t}{t_\bo}\Big)^{-\frac{2(8\mu n +5n +5)}{3(16\mu n + 9n + 9)}} [1+\log(t/t_\bo)]^{\beta} \\[2ex] \propto
&\begin{cases}
t^{-0.36} [1+\log(t/t_\bo)]^{0.03}  , \quad n=3/2\\
\\
t^{-0.36} [1+\log(t/t_\bo)]^{-0.01},\quad n=3
\end{cases}
\end{split}
\end{equation}
where
\begin{equation}
\beta = \frac{4(\mu n +n+1)(2\mu n-1)}{3(\mu n-n-1)(16\mu n +9n+9)} ~.
\end{equation}
%This result is similar the powerlaws found in \cite{ns10}, although the time available for thermalization in a given shell is now $t$ and not $t_\diff$.
When the radiation is in thermal equilibrium , the logarithmic correction is negligible. The reason for that is that in equilibrium, $T_\cl$ can be expressed as:
\begin{equation}
T_\cl = \bigg(\frac{\tau_\cl L}{4 \pi r_\cl^2 a c}\bigg)^{1/4}~.
\end{equation}
As we saw, $L$ and $m_\cl$ depend very weakly on the logarithmic correction, and therefore so do $\tau_\cl$, $r_\cl$ and as a result also $T_\cl$.
%where we omitted the very weak dependence of the result on the logarithmic correction.
\subsection{$T_\obs$ during the transition to the spherical phase} \label{s:T_obs_transition}
Once the breakout shell enters the spherical phase at $t_\s = R/v_\bo$, the energy in the envelope has an inner adiabatic profile ($m>m_\p$) and an outer diffusive profile ($m<m_\p$). At $t_\s<t$, the shell that satisfies $\tau=c/v$, which is initially located at $m_\bo$, starts propagating through the external diffusive profile and increases in mass. 
%
%While crossing the external diffusive profile, the luminosity shell does not generate its own photons, but merely releases the radiation that diffused out of the planar luminsoity shellduring the planar phase.

As explained in Section \ref{s:L_transition}, the transition to the spherical phase starts at $t_\s$ and ends when $\tau=c/v$ reaches $m_\p$.
At the end of the transition phase, the ejecta has no recollection of what happened in the planar phase, in the sense that the increase of $m_\lum$ during the planar phase does not affect the temperature at $t_\tr<t$.
Nevertheless, the logarithmic correction to $m_\lum$ in the planar phase has an effect on the temperature of the transition phase, both when the radiation was thermalized at $t_\s$, and when it was not. We treat these two scenarios in the following subsections.

\subsubsection{$T_\obs$ during the transition phase for $\eta_\lum(t_s)/\xi_\lum(t_s)>1$}  \label{s:T_trans_noneq}
We first consider the case in which the luminosity shell is out of thermal equilibrium at $t_\s$. 
As demonstrated by Eq \eqref{e:eta_planar_diffusive}, during the planar phase, the thermal coupling increases with time, as the value of $\eta$ decreases for a given $m$. During the spherical phase, however, the thermal coupling of a shell becomes weaker, as we show next.

The evolution of $\eta$ is found using equations \eqref{e:rho_sp}, \eqref{e:u_diffusive_spherical}, \eqref{e:u_adiabatic_spherical}, \eqref{e:T_BB} and Eq \eqref{e:eta_ff}:
\begin{equation} \label{e:eta_spherical}
\begin{split}
&\eta(m) = \\
&\begin{cases}
\eta_\bo(t_\bo)\Big(\frac{t_\s}{t_\bo}\Big)^{-1/6} \Big(\frac{t}{t_\s}\Big)^{3/2}\Big(\frac{m}{m_\bo}\Big)^{-\frac{9}{8}-\frac{11\mu n}{3(n+1)}} \\[2ex] \qquad \qquad \qquad\qquad \qquad \qquad ,m(\tau=c/v)<m<m_\p \\[2ex]
\eta(m_\p,t_{\s,\p}) \Big(\frac{t}{t_{\s,\p}}\Big)^{3/2} \Big(\frac{m}{m_\p}\Big)^{-\frac{102\mu n +27n+20}{24(n+1)}}, \quad m_\p<m
\end{cases}
\end{split}
\end{equation}
where $t_{\s,\p}=R/v(m_\p)$ and we neglected for simplicity the weak dependence on the logarithmic factor.
%In the adiabatic energy profile created at $m_\p<m$, $\eta$ is found in the same manner, using Equations \eqref{e:rho_sp} and \eqref{e:u_adiabatic_spherical} to find $T_\BB$:
%\begin{equation}\label{e:eta_spherical_adiabatic}
%\begin{split}
%\eta(m_\p<m) = 
%\eta(m_\p,t_{\s,\p}) \Big(\frac{t}{t_{\s,\p}}\Big)^{3/2} \Big(\frac{m}{m_\p}\Big)^{-\frac{102\mu n +27n+20}{24(n+1)}}  ~.
%\end{split}
%\end{equation}
%where $t_{\s,\p}=R/v(m_\p)$.

Eq \eqref{e:eta_spherical} shows that $\eta$ increases with time, and therefore receives its minimal value at $t=R/v(m)$, which is the time a shell of mass $m$ enters the spherical phase. 
Consequently, shells that were out of themral equilibrium at the end of the planar phase remain so during the spherical phase.

The number of photons in a shell in which $\eta>1$ during the spherical phase is determined by the minimal value of $\eta$ in Eq \eqref{e:eta_planar_diffusive}, namely:
\begin{equation}
\begin{split}
&\eta_\min (m)= \eta(m,t=R/v(m)) = \\[2ex]
&\eta_\bo(t_\bo)\bigg(\frac{t_\s}{t_\bo}\bigg)^{-1/6} \bigg(\frac{m}{m_\bo}\bigg)^{-\frac{52\mu n+27n+27}{24(n+1)}} ~,
\end{split}
\end{equation}
again, neglecting the weak dependence on the logarithmic correction for simplicity.
%The number of photons in a shell in which $\eta>1$ during the spherical phase will be determined by the minimal value of $\eta$, namely:
%\begin{equation}
%\begin{split}
%&\eta_\min (m)= \eta(m,t=R/v) = \\[2ex]
%&\begin{cases}
%\eta_\bo(t_\bo)\Big(\frac{t_\s}{t_\bo}\Big)^{-1/6} \Big(\frac{m}{m_\bo}\Big)^{-\frac{52\mu n+27n+27}{24(n+1)}} \quad  ,m_\bo<m<m_\p \\[2ex]
%\eta(m_\p,t_{\s,\p})\Big(\frac{m}{m_\bo}\Big)^{-\frac{66\mu n+27n+20}{24(n+1)}} \qquad \qquad  ,m_\p<m
%\end{cases}
%\end{split}
%\end{equation}
In the diffusive energy profile at $m<m_\p$, the radiation in each coordinate $m$ has the same characteristic energy (only reduced by adiabatic cooling), and is the typical photon temperature that resided in the shell of mass $m_\p$ at $t_\s$. As the breakout shell enters the spherical phase, the shell that now satisfies $\tau=c/v$ starts propagating inwards in the external energy profile. All shells external to $m_\p$ are also out of thermal equilibrium, since they were so at the end of the planar phase and now their thermal coupling only becomes weaker. Therefore, the shell in which $\tau=c/v$ is out of thermal equilibrium and does not generate its own photons. It merely releases the energy that had reached there by diffusion from inner shells during the planar phase. The temperature of each shell is thus the temperature of the luminosity shell at $t_\s$, which has evolved adiabatically:\begin{equation}
\begin{split}\label{e:T_obs_m_tran_noneq}
&T(m<m_\p,t_\s<t<t_\tr) \\[2ex]
&=T_\obs(t_\s) \bigg(\frac{R/v(m)}{t_\s}\bigg)^{-1/3}\bigg(\frac{t}{R/v(m)}\bigg)^{-1}\\[2ex]
&= T_{\BB,\lum}(t_\s) \frac{\eta_\lum(t_\s)^2}{\xi_\lum(T_\s)^2}
\bigg(\frac{m}{m_\bo}\bigg)^{\frac{2\mu n}{3(n+1)}}\bigg(\frac{t}{t_\s}\bigg)^{-1} ~,
\end{split}
\end{equation}
where $T_\s \equiv T_\obs(t_\s)$.
The observed temperature is determined by the photons that diffuse out of $\tau=c/v$. Sustituting Eq \eqref{e:m_lum_sph} into Eq \eqref{e:T_obs_m_tran_noneq} we find:
\begin{equation}\label{e:T_obs_tran_noneq}
\begin{split}
T_\obs(t_\s<t<t_\tr)=T_{\BB,\lum}(t_\s) \frac{\eta_\lum(t_\s)^2}{\xi_\lum(T_\s)^2} \bigg(\frac{t}{t_\s}\bigg)^{-1 + \frac{4\mu n}{3(\mu n+n+1)}} ~.
\end{split}
\end{equation}
At $t_\tr$, the luminosity shell starts probing shells in which the temperature profile is not constant as before. The luminosity shell might still be out of thermal equilibrium if $\eta_\min(m_\p)/\xi(m_\p)>1$. 
In that case, the observed temperature will drop quickly, until the luminosity shell reaches the shell that satisfies $\eta_\min=1$, which was in thermal equilibrium when it entered the spherical phase. The time when this occurs is denoted by $t_1$ :
\begin{equation}\label{e:t1}
t_1 = t_\s\Bigg[\eta_\bo(t_\bo)\bigg(\frac{t_\bo}{t_\s}\bigg)^{1/6}\Bigg]^{\frac{12(\mu n +n+1)}{66\mu n+27n+20}} ~.
\end{equation} 
At $t_\tr<t<t_1$, the temperature evolves according to:
\begin{equation} \label{e:T_obs_sph_noneq_tr_t_t1}
\begin{split}
&T_\obs(t_\tr<t<t_1) = \frac{\eta_{\min,\lum}^2}{\xi_\lum^2} T_{\BB,\lum}  \\[2ex]
&=T_{\BB,\lum}(t_\tr) \frac{\eta_{\min,\lum}^2(t_\tr)}{\xi_\lum^2(T_\tr)} \xi_\lum^{-2}\bigg(\frac{t}{t_\tr}\bigg)^{-\frac{11+15n+33\mu n}{3(\mu n+n+1)}} ~,
\end{split}
\end{equation}
where $T_\tr \equiv T_\obs(t_\tr)$.
Since thermal coupling is kept through adiabatic cooling when the energy is radiation dominated, the observed radiation will go into thermal equilibrium at $t_1$. 
$\xi_\lum$ decreases with temperature, and therefore $T_\obs$ will not behave as a simple power law, and also in this phase the evoluion of $T_\obs$ is computed numerically.

At $t_1<t$, the observed radiation is in thermal equilibrium, and $T_\obs$ is simply the blackbody temperature of the luminosity shell, computed using equations \eqref{e:rho_sp}, \eqref{e:m_lum_sph}, \eqref{e:u_adiabatic_spherical} and \eqref{e:T_BB}:
\begin{equation}\label{e:T_obs_sph_noneq_t1_t_t2}
\begin{split}
T_\obs(t_1<t<t_2) = T_\obs(t_1)\bigg(\frac{t}{t_1}\bigg)^{-\frac{2+3n}{6(\mu n+n+1)}}
\end{split}
\end{equation}
where $t_2$ is the time where $\eta_\lum = 1$ and the observed temperature from that point is determined in the shell that satisfies $\eta=1$. NS10 have found the colour temperature to evolve roughly as $T_\cl\propto t^{-0.6}$ at $t_2<t$, where the exact value depends on the chioce of $n$.
$\eta_\lum$ is found by substituting Eq \eqref{e:m_lum_sph} into Eq \eqref{e:eta_spherical} for $m_\p<m$, and requiring that $\eta_\lum=1$, we find an expression for $t_2$:
\begin{equation}\label{e:t2}
t_2 = t_\tr \eta(m_\p,t_\tr)^{\frac{12(\mu n+n+1)}{84\mu n+9n+2}} ~.
\end{equation}

%As shown in Section \ref{s:T_obs_planar}, during the planar phase, the thermal coupling increases with time, as the value of $\eta$ decreases for a given $m$.
%During the spherical phase, however, the thermal coupling of a shell decreases with time. 

%\begin{flushright}
%Shells that lie externally to the luminosity shell at the end of the planar phase posses a diffusive energy profile, and show the following evolution of the $\eta$ parameter:
%\begin{equation} \label{e:eta_transition_diffusive}
%\begin{split}
%\eta(m,t) = &\eta_\bo(t_\bo) \Big(\frac{t}{t_\bo}\Big)^{\frac{7}{2} - \frac{7 \mu n}{3(\mu n +n +1)}} \Big(\frac{m}{m_\bo}\Big)^{-\frac{17+17n+28\mu n}{8(n+1)}} \propto  \\[0.90ex] \propto
%&\begin{cases}
%~t^{3.26}~ m^{-2.52}  , \quad n=3/2\\[0.75ex]
%\\
%~t^{3.21}~ m^{-2.62},\quad n=3 ~.
%\end{cases}
%\end{split}
%\end{equation}
%
%When the luminosity shell is in thermal equilibrium, the color temperature is the blackbody temperature of the shell for which $\eta=1$:
%\begin{equation}\label{e:T_cl_trans_equilibrium}
%T_\cl  \propto t^{a} \propto
%\begin{cases}
%~t^{-0.60} , \quad n=3/2\\
%\\
%~t^{-0.60},\quad n=3 ~.
%\end{cases}
%\end{equation}
%where
%\begin{equation}
%a = -\frac{2(15+30n+39\mu n) +n^2(15+39\mu +56\mu^2)}{3(\mu n+n+1)(17+17n+28\mu n)} ~.
%\end{equation}
%The powerlaws in Eq \eqref{e:T_cl_trans_equilibrium} are very similar to the behaviour in the spherical phase (see NS10) due to the very similar energy profiles in the inner adiabatic and external diffusive zones. 
%\end{flushright}
\subsubsection{$T_\obs$ during the transition phase for $\eta_\lum(t_s)/\xi_\lum(t_s)<1$} \label{s:T_trans_eq}
If the luminosity shell is in thermal equilibrium at the end of the planar phase, then all shells with $m<m_\cl(t_\s)$ have the same temperature at $t_\s$, namely the colour temperature. As the shell in which $\tau=c/v$ propagates in $m<m_\cl(t_\s)$, the observed temperature is the adiabatically cooling colour temperature at $t_\s$. 
It is impotrant to note that although the luminosity shell is in thermal equilibrium, the energy release is controlled by the shell that satisfies $\tau=c/v$ which is not necessarily in thermal equilibrium. Therefore, if $m_\bo<m_\cl<m_\p$, initially the radiation is out of thermal equilibrium.

Internal to $m_\cl(t_\s)$, all shells are in thermal equilibrium at $t=t_\s$, and their temperature is their blackbody temperature, $T_\BB(m)$. Since thermal equilibrium is kept through adiabatic cooling, the temperature of these shells continues to be their blackbody temperature.  We denote the time when $m(\tau=c/v)=m_\cl(t_\s)$ by $t_\c$:
\begin{equation}
t_\c = t_\s\bigg(\frac{m_\cl(t_\s)}{m_\bo}\bigg)^{\frac{\mu n+n+1}{2(n+1)}} ~.
\end{equation}
The typical temperature of a shell is:
\begin{equation}
\begin{split}
T(m) =
\begin{cases}
T_\cl(t_\s) \Big(\frac{R/v(m)}{t_\s}\Big)^{-1/3}\Big(\frac{t}{R/v(m)}\Big)^{-1}   \qquad \quad, m<m_\cl(t_\s)\\[2ex]
T_\BB(t_\s) \Big(\frac{R/v(m)}{t_\s}\Big)^{-1/3}\Big(\frac{t}{R/v(m)}\Big)^{-1} \quad  \quad , m_\cl(t_\s)<m
\end{cases}
\end{split}
\end{equation}
As before, the observed temperature is determined by location of $\tau=c/v$:
\begin{equation}\label{e:T_obs_eq_transition}
\begin{split}
&T_\obs = T(\tau=c/v) = \\[1.5ex]
&\begin{cases}
T_\cl(t_\s) \Big(\frac{t}{t_\s}\Big)^{-1 + \frac{4\mu n}{3(\mu n+n+1)}}  \qquad \qquad \qquad  \quad ,t_\s<t<t_\c \\[2ex]
T_\cl(t_\s) \Big(\frac{t_\e}{t_\s}\Big)^{-1 + \frac{4\mu n}{3(\mu n+n+1)}} \Big(\frac{t}{t_\e}\Big)^{\frac{6\mu n-3n-3}{6(\mu n+n+1)}} \quad, t_\e<t<t_{2,\d}
\end{cases}
\end{split}
\end{equation}
The temporal evolution for $t_\s<t<t_\c$ is the same as that in Eq \eqref{e:T_obs_tran_noneq}, since it is simply an adiabatic evoluion of a single temperature.

For some progenitor properties the shell where $\tau=c/v$ can reach $\eta=1$ while still in the transition phase, i.e., while $m_\bo<m(\tau=c/v)<m_\p$. We call this time $t_{2,\d}$, to differ it from $t_2$ in Eq \eqref{e:t2} (which occurs if the luminosity shell reaches $\eta=1$ during the spherical phase, while propagating at $m_\p<m$). 
Starting $t_{2,\d}$, the luminosity shell generates enough photons to reach thermal equilibrium, and the observed temperature is $T_\cl=T_\BB(\eta=1)$.
The time $t_{2,\d}$ is found by substituting $m(\tau=c/v)$ (Eq \ref{e:m_lum_sph}) into Eq \eqref{e:eta_spherical} for $m(\tau=c/v)<m<m_\pl$ to find $\eta(\tau=c/v)$ and requiring $\eta(\tau=c/v)=1$ :
\begin{equation}
t_{2,\d} = t_\s \Bigg[\eta_\bo(t_\bo)\Big(\frac{t_\bo}{t_\s}\Big)^{1/6}\Bigg]^{\frac{12(\mu n+n+1)}{84\mu n+9n+2}}
\end{equation}
Using the specific energy profile at $\tau<c/v$ given by Eq \eqref{e:u_external_transition}, we find the evolution of $\eta$ at $m<m(\tau=c/v)$ 
\begin{equation}\label{e:eta_external_transition}
\begin{split}
&\eta(m<m(\tau=c/v)) \\[2ex]
&=\eta_\bo(t_\bo)\bigg(\frac{t_\s}{t_\bo}\bigg)^{-1/6}\bigg(\frac{m}{m_\bo}\bigg)^{-\frac{17}{8}-\frac{7\mu n}{2(n+1)}} \bigg(\frac{t}{t_\s}\bigg)^{\frac{7}{2}-\frac{7\mu n}{3(\mu n+n+1)}}
\end{split}
\end{equation}
(again, neglecting the weak dependence on the logarithmic correction). The mass of the colour shell is found by requiring $\eta=1$ and using the solution in Eq \eqref{e:eta_external_transition}:
\begin{equation}
\begin{split}
m_\cl = &m_\bo \bigg[\eta_\bo(t_\bo)\bigg(\frac{t_\s}{t_\bo}\bigg)^{-1/6}\bigg]^{\frac{28\mu n+17(n+1)}{8(n+1)}}\\[2ex]
& \times \bigg(\frac{t}{t_\s}\bigg)^{\frac{28(n+1)(\mu n+3n+3)}{3(\mu n+n+1)(28\mu n+17n+17)}} ~.
\end{split}
\end{equation}
The colour temperature during the transition phase is therefore:
\begin{equation}
T_\cl = T_\obs(t_{2,\d}) \bigg(\frac{t}{t_{2,\d}}\bigg)^{\chi}~,
\end{equation}
where
\begin{equation}
\chi = -\frac{2(15+30n+39\mu n)+n^2(15+39\mu +56 \mu^2)}{3(\mu n+n+1)(28\mu n+17n+17)} \approx -0.60
\end{equation}
for both {$\mu=0.19,n=3/2$} and {$\mu=0.19, n=3$}.

\section{Light Travel Time Effects}

In a spherical explosion, radiation emitted from different parts of the progenitor surface arrives at the observer at different times. Due to this effect, the observed radiation at time $t$ will be a mix of light emitted at different times, and thus at different intensities.
This effect is important if $t_\bo<R/c$ and affects the observed radiation at $t<R/c$. The apparent luminosity, $L_{\text{ltt}}$, is calculated using the following integral:
\begin{equation}
\begin{split}
L_{\text{ltt}}(t) &= \int^{R}_{max\{0,R-ct\}}L(t') \frac{r}{R^2}dr \\
&=\int^{t}_{max\{0,t-R/c\}} \frac{L(t')}{R/c}\bigg[1-\frac{c}{R}(t-t')\bigg] dt' ~,
\end{split}
\end{equation}
where $r$ is the projected radius and $t' = t-\Delta t$ is the time at the emitting source, where $\Delta t = (R-\sqrt{R^2-r^2})/c$ is the travel time of photons emitted from apparent radius $r$ to the plane tangent to the front of the sphere.
The luminosity is thus smeared over a timescale of $\sim R/c$ and for $t\ll R/c$ it is roughly constant and equal to $\sim L_\bo t_\bo c/R$. At $R/c<t$ light travel time ceases to affect the observed luminosity and $L\propto t^{-4/3}$ as in Eq \eqref{e:L_planar}.

The observed spectrum is also affected by the different arrival times of the emitted radiation. At $t<t_\bo$ the spectrum is dominated by radiation escaping from the shock front, and the typical observed photon energy will be $T_\bo$. While $t_\bo<t<R/c$, the observed spectrum will be a combination of radiation arriving from different apparent radii, each with a different typical energy $T$, and thus will not be a simple blackbody (or diluted blackboody) spectrum.
The observed spectrum receives a simple form if the intrinsic temperature and luminosity evolve as a power laws. This is a very good approximation for the luminosity, for which the logarithmic correction has no significant effect, but does not hold for the temperature, as evident from Eq \eqref{e:T_obs_planar}.
Nevertheless, the early temperature evolution, including the logarithmic correction, can be fit relatively well by a power law. Taking $L(t) = L_\bo(t/t_\bo)^{-4/3}$ and $T(t) = T_\bo(t/t_\bo)^{-\alpha}$, the spectrum broadens in time to form a power law:

\begin{equation} \label{e:F_nu_ltt}
F_\nu \propto \frac{d L}{d \nu} \propto \nu^{\frac{1}{3\alpha} -1}~.
\end{equation}
We will show in the next section that typically $1/(3\alpha)>0$, so that $\nu F_\nu$ is an increasing function of $\nu$. The upper end of this spectrum corresponds to the initial breakout temperature, $T_\obs(t_\bo)$, and the lowest observed frequency corresponds to the non-delayed temperature, $T_\obs(t)$. $\nu F_\nu$ is a slowly increasing function of $\nu$, and therefore no typical energy can be defined during this stage. The observed spectrum will thus be broad band until $t=R/c$.
%Assuming $L(t'<t_\bo) = L_0$ and using $L(t_\bo<t'<t_\s) = L_\bo(t/t_\bo)^{-4/3}$ (ignoring the logarithmic correction for the luminosity), we find:
%
%\begin{equation}
%L_{\text{ltt}}(t) = \frac{L_\bo ~t_\bo}{R/c}\bigg\{1-\frac{t}{R/c}+3 \bigg[1-\Big(\frac{t}{t_\bo}\Big)^{-1/3}\Big(\frac{5}{2}\frac{t}{R/c}+1\Big)\bigg]\bigg\}
%\end{equation}

\section{Applications to various progenitors} \label{s:progenitors}
In this section we apply our results to different progenitors: RSG, BSG and WR, and discuss the effect of the planar logarithmic correction on the observed properties of each progenitor.

The expressions derived in the previous sections for $T_\cl, T_\obs$ and $L$ are functions of the progenitor density structure and shock velocity, which depend on the progenior peoperties $E, M, R$ and $\kappa$, on the dimensionless conctants $C_1$ and $C_2$ defined in Eq \eqref{e:kappa} and on the density power law index $n$. Throughout the paper we assume a single power law density profile. This assumption may not accurately describe the profile of a realistic progenitor in the entire radii range relevant for this analysis \citep[e.g.,][]{Morozova16}. However, the model is not very sensitive to the exact power law index, but rather depends more on the normalization of the profiles.

The constants $C_1$ and $C_2$ determine the normalization of the envelope structure and velocity, when the density is described by a polytrope. We use \cite{matzner&mckee} to estimate these conatants, which depend on whether the enveope is radiative of convective. For each of the progenitors discussed here we provide the apropriate values of $C_1$ and $C_2$.

We use the following notations to specify the progenitor properties:
$M_x = x M_\odot$, $R_x = x R_\odot$, $E_x = 10^x \text{erg}$ and $\kappa_x = x~ \text{cm}^2 \text{g}^{-1}$.

\subsection{Red Supergiant}
RSGs are the progeniors of most type-II explosions. They have a typical radius of $500 R_\odot$ and since their envelopes are hydrogen rich, their typical Thomson opacity is $\sim 0.34 \text{ cm}^2 \text{g}^{-1}$. The envelopes of RSGs are convective, and therefore $n=3/2$.
If we assume that the RSG ejecta is a polytrope all the way to the core, we find that $C_1\sim 1$ and $C_2\sim 0.8$. Since this is a rather simplified assumption, we use instead equation 14 in \cite{matzner&mckee} to estimate these constants, which now depend on the detailed envelope structure.
Taking typical values for the total, ejecta and envelope mass of $M_* = 15 M_\odot$, $M_\ej =13M_\odot$ and $M_{\text{env}}= 10M_\odot$, respectively, we find $C_1\sim 0.5$ and $C_2\sim 0.9$. The values of $C_1$ and $C_2$ fit well the numerical profiles of a progenitor with similar properties, evolved using MESA \citep{Paxton11}.

For the above typical RSG properties, the breakout time according to Eq \eqref{e:ti_bo} is:
\begin{equation}\label{e:tbo_RSG}
t_\bo^{\text{RSG}} = ~92 \text{ s} ~  M_{15}^{0.21} R_{500}^{2.16} E_{51}^{-0.79} \kappa_{0.34} ^{-0.58}
\end{equation}
The breakout shell enters the spherical phase at:
\begin{equation}\label{e:ts_RSG}
t_\s = 7.2 \text{ hr } M_{15}^{0.44} R_{500}^{1.26} E_{51}^{-0.56} \kappa_{0.34}^{-0.13} ~,
\end{equation}
according to Eq \eqref{e:t_s}.
During the planar phase, the luminosity evolves according to Eq \eqref{e:L_planar}, whereas during the transition from the planar to the spherical phase, we estimate the luminosity using Eq \eqref{e:L_transition}:
\begin{equation}\label{e:L_RSG}
\begin{split}
&L^{\text{RSG}}(t) = \\[2ex]
&\begin{cases}
&5.5\times 10^{45}  \text{ erg}\cdot \text{s}^{-1} M_{15}^{-0.37} R_{500}^{2.46} E_{51}^{0.30} \kappa_{0.34}^{-1.06} \\[1.7ex]
&\times[1+ \log(t/92~\text{s})]^{0.06} \Big(\frac{t}{90~\text{s}}\Big)^{-4/3} \quad \quad  ,t_\bo<t<t_\s \\[2ex]
&5.8\times10^{41}  \text{ erg}\cdot \text{s}^{-1} M_{15}^{-0.83} R_{500}^{1.13} E_{51}^{0.90} \kappa_{0.34}^{-0.93} \\[2ex]
&\times\Big(\frac{t}{1\text{hr}}\Big)^{-0.27}  \qquad \qquad \qquad \qquad  \quad,t_\s<t<t_\tr
\end{cases}
\end{split}
\end{equation}
where $t_\tr = 24 ~\text{hr}$ according to Eq \eqref{e:t_transition} and the estimates for $t_\bo$ and $t_\s$ obtained in Eq \eqref{e:tbo_RSG} and \eqref{e:ts_RSG}.

%The evolution of the luminosity at $t_\s<t<t_\tr$ is not very different from the behaviour during the spherical phase (e.g., \citealt{ns10}). The reason is that the energy density profile varies very little between the internal and external parts of the luminosity shell. For example, during the transition phase the radiation leaks out of regions with an energy density of $\varepsilon \propto m^{1.23}$, whereas during the spherical phase the photons leak out of regions with $\varepsilon \propto m^{1.46}$ (for $n=3/2$). Therefore, the luminosity is only marginally affected by the transition between the two density profiles.

\begin{figure}
 \centering
\includegraphics[width=0.9\columnwidth]{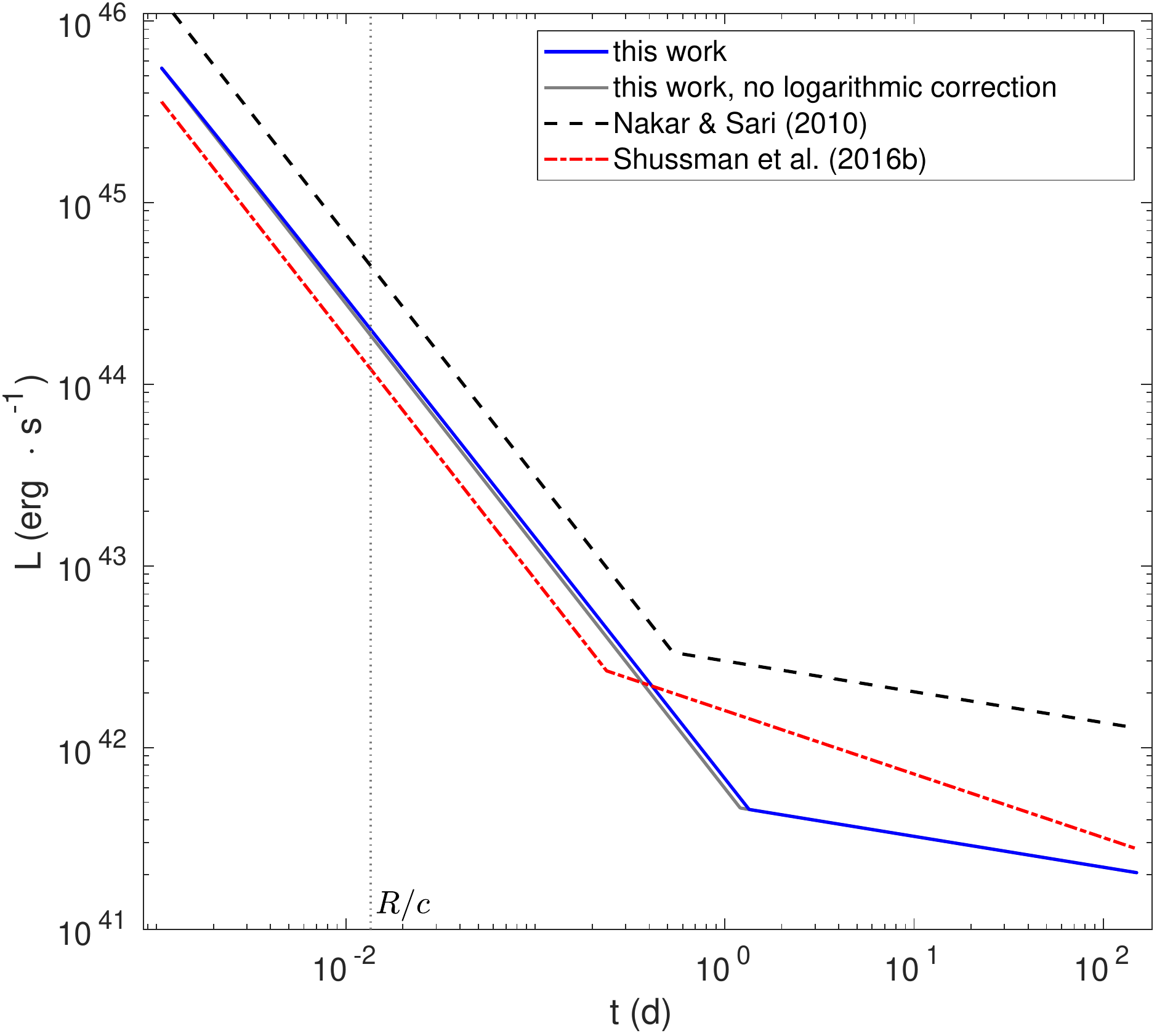} 
\caption{The intrinsic bolometric luminosity of an RSG with the following parameters: $M_{15}=1$, $R_{500}=1$,$E_{51}=1$ and $\kappa_{0.34}=1$. Our results are compared to previous analytic works that consider a similar external density profile for the progenitor star. The break in the light curve indictes the transition to the spherical phase. The observed bolometric luminosity will be smeared on a time scale of $R/c\sim 1200 \text{ s}$ due to light travel time effects.}\label{f:L_RSG}
\end{figure}

In Figure \ref{f:L_RSG} we show the luminosity calculated in this work (excluding the transition phase) with and without the logarithmic correction, together with those of NS10 and \cite{Shussman2016b} for $n=3/2$ for the typical RSG properties. \cite{Shussman2016b} perform radiative transfer calculations on a set of analytic progenitors with a powerlaw density profile, and calibrate the numerical factors of the analytic model using the simulated results. The luminosity of NS10's model is a factor of $\sim 2$ higher than the luminosity calculated in this work, while the numerically calibrated luminosity of \cite{Shussman2016b} is $\sim 1.6$ times lower. The behaviour of the luminosity during the spherical phase is different in \cite{Shussman2016b}'s model since they take into account the transition in the density profile in the inner regions probed during the spherical phase.
From Figure \ref{f:L_RSG} it is evident that the logarithmic correction does not play a role in shaping the observed luminosity, in agreement with Eq \eqref{e:L_planar}.

The behaviour of the temperature during the planar phase depends upon whether the luminosity shell is in thermal equilibrium at $t_\bo$, which is determined by equations \eqref{e:xi_def} and \eqref{e:eta_planar_lum}. For an RSG using Eq \eqref{e:eta_planar_lum}, we find the thermal coupling coefficient during the planar phase:
\begin{equation} \label{e:eta_lum_planar_RSG}
\begin{split}
\eta_\lum(t) =& 1.3~ M_{15}^{-1.68} R_{500}^{-0.40} E_{51}^{2.03}\kappa_{0.34}^{1.47} \bigg(\frac{t}{92~\text{s}}\bigg)^{-1/6} \times \\ & [1+\log(t/92~\text{s})]^{-1.47} ~.
\end{split}
\end{equation}
Hence, the luminosity shell is marginally in thermal equilibrium at breakout time.
In RSGs, Comptonization does not play an important role in thermalizing the radiation, since $k T_\obs$ is not much higher than $h \nu_\min$. Indeed, we find that at breakout time $\xi_\lum(t_\bo)\sim 1$ while $\eta_\lum(t_\bo) \sim 1.3$. This implies that $\eta_\lum/ \xi_\lum >1$ and initially, the radiation is out of thermal equilibrium. However, as the luminosity shell retreats into regions of higher density, the observed radiation becomes thermalized, and $T_\obs$ quickly reaches $T_{\BB,\lum}$. We denote this time as $t_\eq$.
After reaching thermal equilibrium, the colour temperature is determined in the shell where $\eta=1$, i.e., $T_\obs = T_\cl = T_\BB(\eta = 1)$, and $T_{\BB,\lum}>T_\cl$.
At $t_\s$, the SN enters into the transition phase, and the temperature behaves as described in Section \ref{s:T_trans_eq}. At the end of the transition phase, the envelope enters the spherical phase and none of the observed properties are affected by the planar phase logarithmic correction.

%We find that radiation from most RSG explosions will be out of thermal equilibrium at breakout. If during the planar phase the luminosity shell reaches thermal equilibrium, the behavious $T_\obs$ will be:
%\begin{equation}\label{e:T_obs_eq_RSG}
%\begin{split}
%&T_\obs^{\text{RSG}}= \\[1.5ex]
%&\begin{cases}
%&1.5 \times 10^{6} \text{ K}~  M_{15}^{-3.35} R_{500}^{-0.62} E_{51}^{3.98}\kappa_{34}^{2.64} \Big(\frac{\xi_\lum}{1.8}\Big)^{-2}  \\[1.3ex]
%&\times \Big(\frac{t}{54~\text{s}}\Big)^{-2/3}[1+log(t/t_\bo)]^{-2.64} \\[2ex]   
%& \qquad \qquad \qquad \qquad \qquad \qquad \qquad \qquad,t_\bo<t<t_\eq \\[2ex]
%&~2.9 \times 10^{5}\text{ K}~  M_{15}^{-0.29} R_{500}^{0.11} E_{51}^{0.31}\kappa_{34}^{-0.03} \Big(\frac{t}{300 \text{s}}\Big)^{-0.36} \\[1.3ex] & \times[1+log(t/t_\bo)]^{0.03} \qquad \qquad \qquad \quad, t_\eq<t<t_{\s,\bo} \\[2ex]
%&~6.4 \times 10^{4}~ \text{ K}~  M_{15}^{-0.08} R_{500}^{0.73} E_{51}^{0.03}\kappa_{34}^{-0.09}\Big(\frac{t}{\text{6~hr}}\Big)^{-0.86} \\
%& \qquad \qquad \qquad \qquad \qquad \qquad \qquad \qquad,t_{\s,\bo}<t<t_\c \\[2ex]
%&~4.7 \times 10^{4}~ \text{ K}~  M_{15}^{0.05} R_{500}^{0.28} E_{51}^{-0.11}\kappa_{34}^{-0.31}\Big(\frac{t}{\text{8~hr}}\Big)^{-0.42}\\
%& \qquad \qquad \qquad \qquad \qquad \qquad \qquad \qquad ,t_\c<t<t_\tr
%\end{cases}
%\end{split}
%\end{equation}
We use equations \eqref{e:T_obs_planar}, \eqref{e:T_cl_planar} and \eqref{e:T_obs_eq_transition} to obtain:
\begin{equation}\label{e:T_obs_eq_RSG}
\begin{split}
&T_\obs^{\text{RSG}}= \\[1.5ex]
&\begin{cases}
&5.3 \times 10^5 \text{ K}~  M_{15}^{-3.35} R_{500}^{-0.62} E_{51}^{3.98}\kappa_{0.34}^{2.64} \Big(\frac{t}{100 \text{s}}\Big)^{-2/3} \\[1.3ex]
&\times [1+\log(t/92 ~\text{s})]^{-2.64}  \qquad \qquad \qquad,t_\bo<t<t_\eq \\[2ex]
&~4.1 \times 10^{5}\text{ K}~  M_{15}^{-0.29} R_{500}^{0.11} E_{51}^{0.31}\kappa_{0.34}^{-0.03} \Big(\frac{t}{100 \text{s}}\Big)^{-0.36} \\[1.3ex] & \times[1+\log(t/92~ \text{s})]^{0.03} \qquad \qquad \qquad \quad, t_\eq<t<t_\s \\[2ex]
&~1.8 \times 10^{5}~ \text{ K}~  M_{15}^{-0.12} R_{500}^{0.42} E_{51}^{0.09}\kappa_{0.34}^{-0.12}\Big(\frac{t}{\text{1~hr}}\Big)^{-0.60}\\
& \qquad \qquad \qquad \qquad \qquad \qquad \qquad \qquad ,t_\s<t<t_\tr
\end{cases}
\end{split}
\end{equation}
where the expression for $t_\bo<t<t_\eq$ is for the non-thermal radiation.
For the above progenitor properties, the breakout shell itself (and therefore also the observed radiation) is in thermal equilibrium at $t_\s$. The time $t_\c$ is not defined for this specific progenitor, since the colour shell at $t_\s$ is external to the breakout shell.

\begin{figure*}
\centering
\includegraphics[width=1.95\columnwidth]{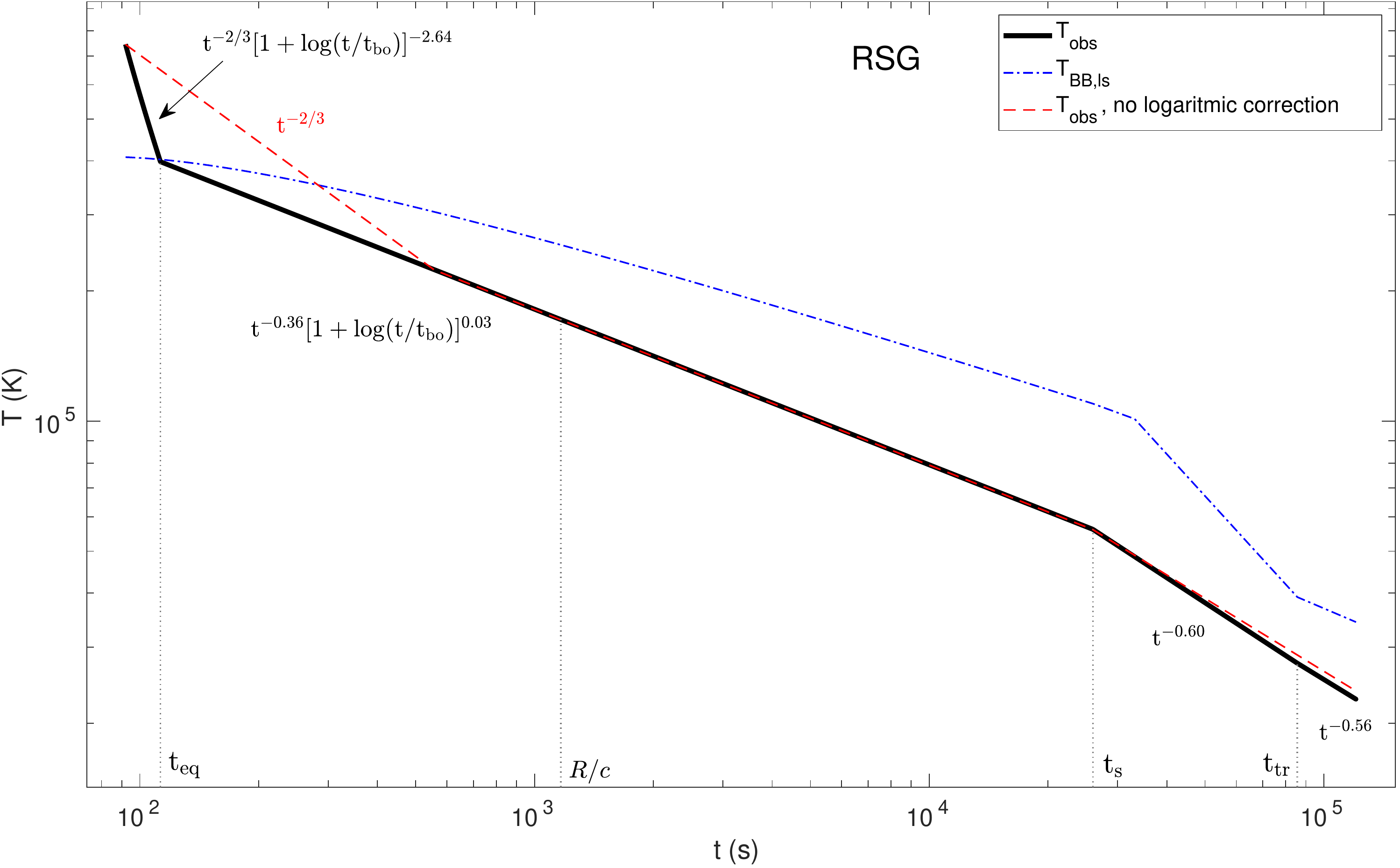} 
\caption{Solid black line: the observed temperature of an RSG with $M_{15}=1$, $R_{500}=1$, $E_{51}=1$ and $\kappa _{0.34}=1$. The dash-dotted blue line is the blackbody temperature of the luminosity shell, and the dashed red line is the observed temperature that would haved been obtained without the logarithmic correction to the luminosity shell. In this plot we show the typical radiation temperature in the progenitor frame (not taking into account light travel time effects). Initially, the radiation is out of thermal equilibrium, as $T_\obs>T_\BB$. Due to the increase of $m_\lum$, the radiation quickly reaches thermal equilibrium at $t=t_\eq$, while the uncorrected $T_\obs$ remains out of thermal equilibrium for a few more minutes before reaching thermalization. While $t<R/c\sim 1200 \text{ s}$, light travel time variations create a broad band spectrum that follows $F_\nu\propto \nu^{-0.07}$, such that no typical temperature can be defined during that period in the observer frame. At $t_\eq<t<t_\s$ the observed temperature is determined in the shell where $\eta=1$. The luminosity enters the transition phase at $t_\s$ and decreases gradually (see text for details). At $t=t_\tr$ the transition phase ends, and the observed temperature is no longer affected by the logarithmic correction.} \label{f:T_obs_RSG}
\end{figure*}

We plot the evolution of $T_\obs$ according to Eq \eqref{e:T_obs_eq_RSG} in Figure \ref{f:T_obs_RSG}, together with the observed temperature derived without the logarithmic correction for comparison, and the blackbody temperature of the luminosity shell.

We find that without the logarithmic correction, the radiation remains out of thermal equilibrium for a few minutes, but thermalizes while still in the planar phase. From that point, $T_\obs<T_{\BB,\lum}$, since the radiation thermalizes in $m_\cl<m_\lum$, where $T_\BB$ is lower than that of the luminosity shell.
This result is only in mild contradiction with previous works (e.g., NS10) that predicted that an RSG will be in thermal equilibrium at breakout and throughout the planar phase, while considering a constant luminosity shell.
For the progenitor properties cosidered, the radiation peaks in the near UV during the planar phase.

During $t<R/c\sim 1200 \text{s}$, the observed spectrum is affected by light travel time variations. Since the radiation is in thermal equilibrium throughout most of this phase, the effective power law that the temperature follows is $T\propto t^{-0.36}$. Therefore, according to Eq \eqref{e:F_nu_ltt} the observed spectrum will follow $F_\nu \propto \nu^{-0.07}$ below the peak of $\nu F_\nu$ at $T_\obs(t_\bo)$.

\begin{figure*}
\centering
\includegraphics[width=1.95\columnwidth]{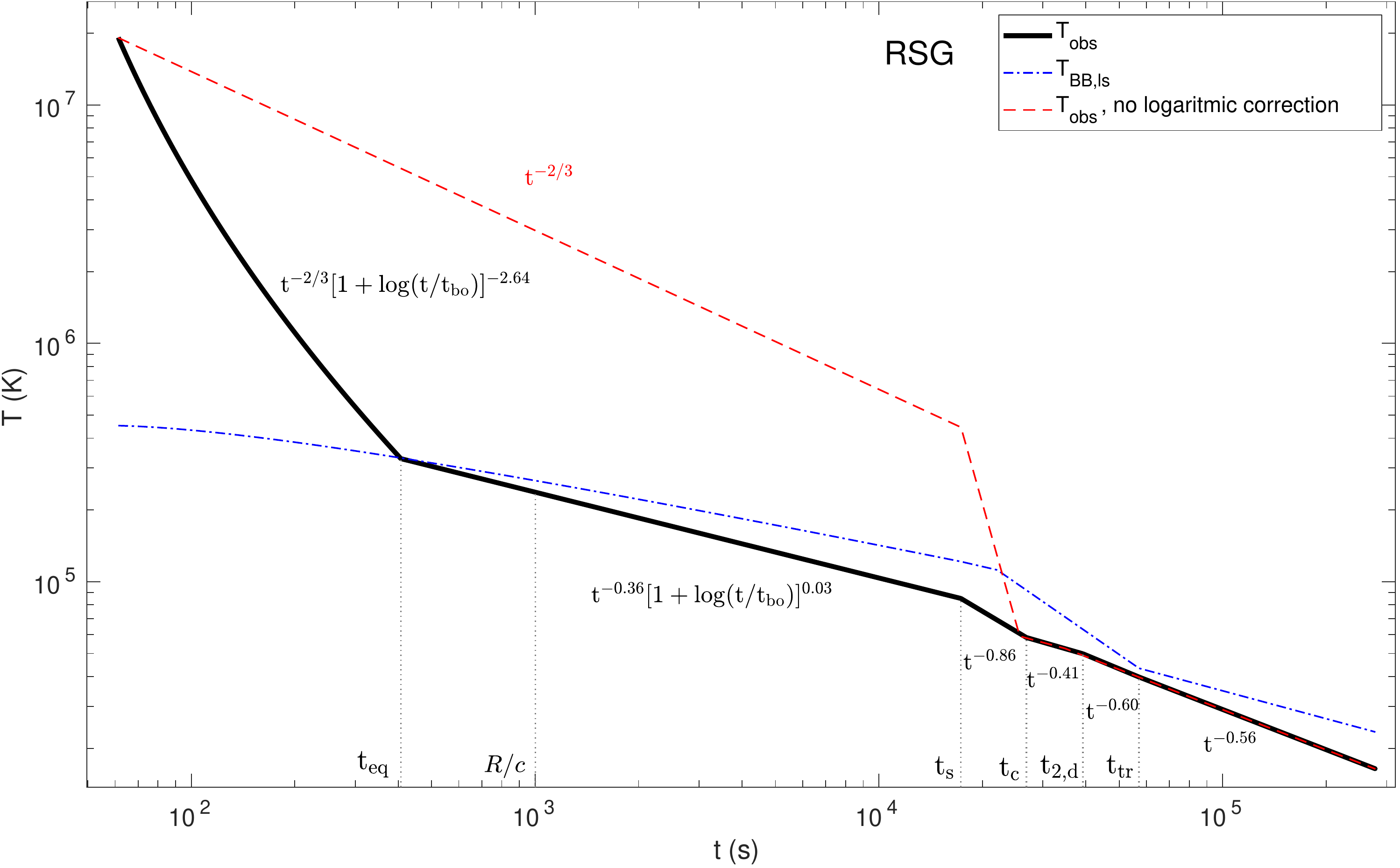} 
\caption{Same as Figure \ref{f:T_obs_RSG} for an RSG with $M_{10}=1$, $R_{500}=1$, $E_{51}=1.5$ and $\kappa _{0.34}=1$. In this case, the spectrum in the observer frame at $t<R/c \sim 1200 \text{ s}$ will follow $F_\nu\propto \nu^{-0.87}$, where the peak of $\nu F_\nu$ is at $T_\obs(t_\bo)$.} \label{f:T_obs_RSG_noneq}
\end{figure*}

For a less massive and more energetic progenitor of e.g. $M=10M_\odot$ and $E = 1.5\times 10^{51}\text{ erg}$, thermal coupling becomes weaker and $\eta_\lum(t_\bo)=6$, in accordance with Eq \eqref{e:eta_lum_planar_RSG}. Without the logarithmic correction, the radiation would have remained out of thermal equilibrium throughout all of the planar phase (see Figure \ref{f:T_obs_RSG_noneq}). However, with the logarithmic correction the radiation reaches thermal equilibrium after $\sim 400$ seconds. 
The breakout and spherical times for these  progenitor properties are:
\begin{equation}\label{e:tbo_RSG_noneq}
%\begin{split}
t_\bo^{\text{RSG}} = ~60 \text{ s} ~  M_{10}^{0.21} R_{500}^{2.16} E_{51.2}^{-0.79} \kappa_{0.34} ^{-0.58}
\end{equation}
and
\begin{equation}\label{e:ts_RSG_noneq}
t_\s = 4.8 \text{ hr}~ M_{10}^{0.44} R_{500}^{1.26} E_{51.2}^{-0.56} \kappa_{0.34}^{-0.13} ~,
\end{equation}

Using equations \eqref{e:T_obs_planar}, \eqref{e:T_obs_tran_noneq}, \eqref{e:T_obs_sph_noneq_tr_t_t1} and \eqref{e:T_obs_sph_noneq_t1_t_t2}, we find the temperature evolution for this case:
\begin{equation}\label{e:T_obs_noneq_RSG}
\begin{split}
&T_\obs^{\text{RSG}}= \\[1.5ex]
&\begin{cases}
&1.9 \times 10^7 \text{ K}~  M_{10}^{-3.35} R_{500}^{-0.62} E_{51.2}^{3.98}\kappa_{0.34}^{2.64} \Big(\frac{t}{60 \text{s}}\Big)^{-2/3} \\[1.3ex]
&\times [1+\log(t/60 ~\text{s})]^{-2.64}  \qquad \qquad \qquad,t_\bo<t<t_\eq \\[2ex]
&~2.8 \times 10^{5}\text{ K}~  M_{10}^{-0.29} R_{500}^{0.11} E_{51.2}^{0.31}\kappa_{0.34}^{-0.03} \Big(\frac{t}{10 \text{m}}\Big)^{-0.36} \\[1.3ex] & \times[1+\log(t/60~ \text{s})]^{0.03} \qquad \qquad \qquad \quad, t_\eq<t<t_\s \\[2ex]
&~3.3 \times 10^{5}~ \text{ K}~  M_{10}^{-0.08} R_{500}^{0.73} E_{51.2}^{0.03}\kappa_{34}^{-0.09}\Big(\frac{t}{\text{1~hr}}\Big)^{-0.86} \\
& \qquad \qquad \qquad \qquad \qquad \qquad \qquad \qquad,t_\s<t<t_\c \\[2ex]
&~5.2 \times 10^{4}~ \text{ K}~  M_{10}^{0.05} R_{500}^{0.28} E_{51.2}^{-0.11}\kappa_{0.34}^{-0.31}\Big(\frac{t}{\text{10~hr}}\Big)^{-0.41}\\
& \qquad \qquad \qquad \qquad \qquad \qquad \qquad \qquad ,t_\c<t<t_{2,\d}\\[2ex]
&~5.4 \times 10^{4}~ \text{ K}~  M_{10}^{-0.12} R_{500}^{0.42} E_{51.2}^{0.09}\kappa_{0.34}^{-0.12}\Big(\frac{t}{\text{10~hr}}\Big)^{-0.60}\\
& \qquad \qquad \qquad \qquad \qquad \qquad \qquad \qquad ,t_{2,\d}<t<t_\tr
\end{cases}
\end{split}
\end{equation}
where
\begin{equation}
t_\c = 7~\text{hr}~ M_{10}^{-0.28} R_{500}^{1.00} E_{51.2}^{0.31} \kappa_{0.34}^{0.49},
\end{equation}
\begin{equation}
t_{2,\d} = 11~\text{hr}~ M_{10}^{-0.95} R_{500}^{0.79} E_{51.2}^{1.10} \kappa_{0.34}^{1.05} ~,
\end{equation}
and $t_\tr \sim 16$ hr. This temperature evolution is plotted in Figure \ref{f:T_obs_RSG_noneq}.

For these progenitor parameters, the radiation first peaks in the X-ray range, but shifts to the UV during the planar phase owing to the logarithmic correction.

A fit to the temperature at the early phases shows that it decreases approximately as $t^{-2.5}$ while it is out of thermal equilibrium. As a result, while the observed spectrum is affected by light travel time varitions (at $t<R/c$), it will follow $F_\nu \propto \nu^{-0.87}$.

\subsection{Blue Supergiant}
Explosions of BSGs also produce type-II SNe (e.g., SN1987A), and are therefore hydrogen rich with $\kappa = 0.34$ cm$^2~$g$^{-1}$. Their typical radius is $R\sim 50 R_\odot$ and have a mass similar to an RSG of $M\sim 15 M_\odot$. However, their envelopes are radiative and therefore $n=3$.
For radiative envelopes we find $C_1\sim 0.35$ and $C_2\sim 0.97$. These values agree with the BSG pre-explosion density profiles and the post explosion velocity profiles in \cite{Dessart18}. We use these values to describe a typical BSG.
%We apply the observed parameters of SN1987A, whose progenitor was a BSG, to estimate the values of $C_1$ and $C_2$. For  $L_* \sim 1.1\times 10^5 L_\odot$ and $M\sim 11 M_\odot$ (see e.g.,\citealt{Woosley1988}, \citealt{Podsiadlowski1992}), we find $C_1\sim 0.35$ and $C_2\sim 0.97$. We use these values to describe a typical BSG.

\begin{figure}
\includegraphics[width=0.9\columnwidth]{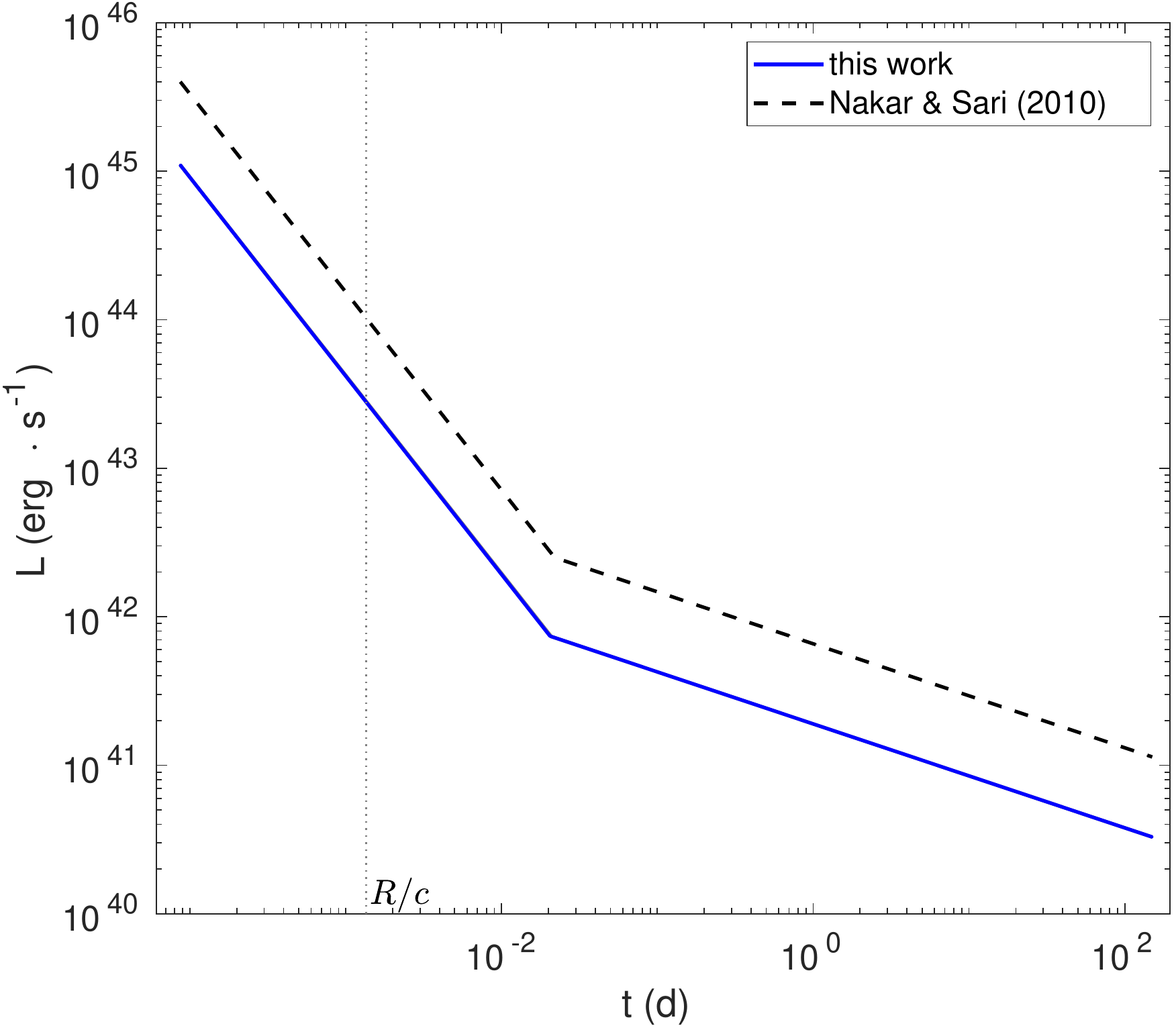} 
\caption{The intrinsic bolometric luminosity of a BSG with the parameters $M_{15}=1$, $R_{50}=1$,$E_{51}=1$ and $\kappa_{0.34}=1$. The observed bolometric luminosity will be smeared on a time scale of $R/c\sim 120 \text{ s}$ due to light travel time effects.}\label{f:L_BSG}
\end{figure}

The breakout time of a BSG is significantly shorter compared to an RSG, since the material is ejected at higher velocities due to the smaller radius and steeper density profile:
\begin{equation}
t_\bo^{\text{BSG}} =  ~7.5 \text{ s }  M_{15}^{0.27} R_{50}^{1.92} E_{51}^{-0.73} \kappa_{0.34} ^{-0.46}  ~.
\end{equation}
For the assumed BSG properties, the planar phase of the breakout shell ends at:
\begin{equation}
t_\s^{\text{BSG}} = ~15 \text{ m } M_{15}^{0.42} R_{50}^{1.33} E_{51}^{-0.58} \kappa_{0.34}^{-0.17} ~,
\end{equation}
and therefore the planar phase evolution of a BSG is not likely to be observed, as the cadence of current surveys is days to hours.
The luminosity of a typical BSG in the planar and the transition phases is obtained using equations \eqref{e:L_planar} and \eqref{e:L_transition}:
\begin{equation} \label{e:L_BSG}
\begin{split}
&L^{\text{BSG}}(t) = \\
&\begin{cases}
&~7.7\times 10^{44}  \text{ erg}\cdot \text{s}^{-1} M_{15}^{-0.33} R_{50}^{2.31} E_{51}^{0.34} \kappa_{0.34}^{-0.99}\times \\
&[1+ \log(t/7~ \text{s})]^{-0.01} \Big(\frac{t}{10~ \text{s}}\Big)^{-4/3} \quad \quad ,t<t_\s \\[2ex]
&1.8\times 10^{41}  \text{ erg}\cdot \text{s}^{-1} M_{15}^{-0.74} R_{50}^{0.97} E_{51}^{0.92} \kappa_{0.34}^{-0.82}\times \\[2ex]
& \Big(\frac{t}{1\text{hr}}\Big)^{-0.33}  \qquad \qquad \qquad \qquad \qquad ,t_\s<t<t_\tr
\end{cases}
\end{split}
\end{equation}
where $t_\tr=47~ \text{m}$. In Figure \ref{f:L_BSG} we plot the luminosity derived in this work along with that obtained by NS10. The result of NS10 is $\sim 3$ higher than what we obtain in Eq \eqref{e:L_BSG}. Radiation from a BSG will be much fainter than that of an RSG  (Eq \ref{e:L_RSG}). The low luminosity is expected for compact progenitors that lose most of their energy to adiabatic expansion, and is expressed by the high power of $R$ in Eq \eqref{e:L_BSG}. A good example is SN1987A, whose light curve is powered mostly by $^{56}$Ni, and has only a small and early contribution from shock cooling radiation \citep[e.g.,][]{Woosley1988}.

A BSG in the planar phase will initially be out of thermal equilibrium with the following value of $\eta_\lum$:
\begin{equation} \label{e:eta_lum_planar_BSG}
\begin{split}
\eta_\lum^{\text{BSG}}(t) =& ~70 ~ M_{15}^{-1.59} R_{50}^{-0.77} E_{51}^{2.12}\kappa_{0.34}^{1.66} \bigg(\frac{t}{7~\text{s}}\bigg)^{-1/6} \\ & \times [1+\log(t/7~ \text{s})]^{-1.66} ~.
\end{split}
\end{equation}
Nevertheless, in the case of BSG explosions, upscattering of lower energy photons by Compton interactions is important due to the high gas temperature, and therefore contributes greatly to thermalization. At breakout, we find that $\xi_\lum\sim 3$, such that $\eta_\lum/\xi_\lum \sim 20$ and the observed radiation is still out of thermal equilibrium. The dependence of $T_\obs$ on time differs from a power law, since $\xi_\lum$ depends logarithmically on $T_\obs$, which itself decreases non-negligibly. We find that at the end of the planar phase, $\eta_\lum\sim 1.7$ and $\xi_\lum\sim 1$, and the radiation is almost in thermal equilibrium. 

\begin{figure*}
 \centering
\includegraphics[width=1.95\columnwidth]{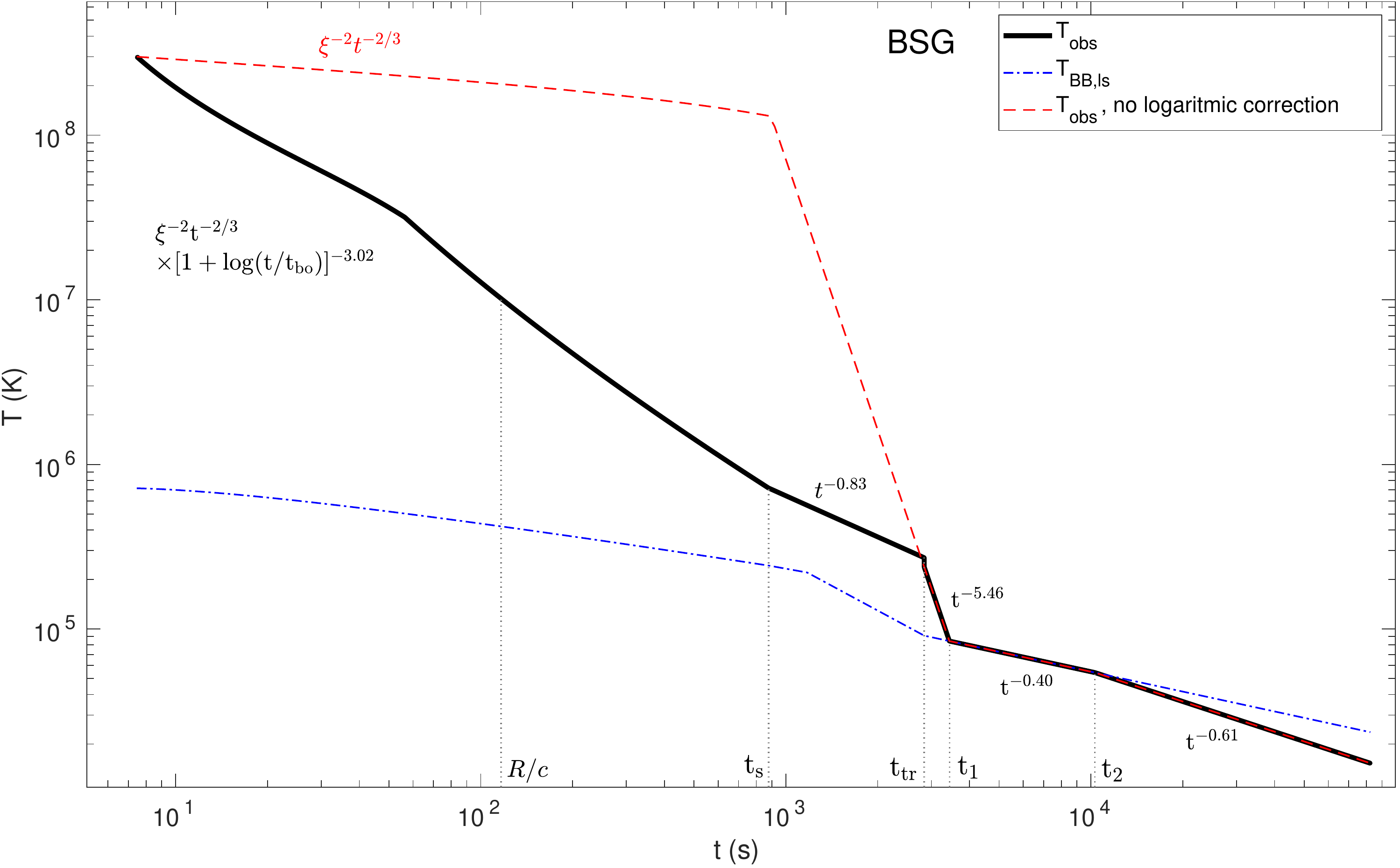} 
\caption{Aame as Figure \ref{f:T_obs_RSG} for a BSG with the parameters $M_{15}=1$, $R_{50}=1$, $E_{51}=1$ and $\kappa _{0.34}=1$. Both the corrected and uncorrected $T_\obs$ remain out of thermal equilibrium throughout the planar phase, but the logarithmic correction causes the observed temperature to decrease significantly during that time, shifting the observed radiation to peak in the UV at $t_\s$, instead of in the X-rays. Light travel time variations create a broad band sectrum in the observer frame at $t<R/c\sim 120 \text{ s}$, which follows $F_\nu\propto \nu^{-0.72}$. The radiation enters the transition phase at $t_\s$, where the luminosity shell probes the external diffusive energy profile. At $t_\tr$, the luminosity shell reaches $m_\p$ and the evolution becomes identical to that of the uncorrected temperature. From that point, $T_\obs$ falls quickly into thermal equilibrium at $t_1$. Due to the faster evolution during the planar phase, the drop in $T_\obs$ between $t_\s$ and $t_1$ is much less dramatic than that obtained without the logarithmic correction.} \label{f:T_obs_BSG}
\end{figure*}

We use equations \eqref{e:T_obs_planar}, \eqref{e:T_obs_tran_noneq}, \eqref{e:T_obs_sph_noneq_tr_t_t1} and \eqref{e:T_obs_sph_noneq_t1_t_t2} to obtain:
\begin{equation}\label{e:T_obs_noneq_BSG}
\begin{split}
&T_\obs^{\text{BSG}}(\eta_\lum>1)= \\[1.5ex]
&\begin{cases}
&2 \times 10^{8} \text{ K}~  M_{15}^{-3.02} R_{50}^{-1.38} E_{51}^{4.18}  \kappa_{0.34}^{3.02} \Big(\frac{\xi_\lum}{2.6}\Big)^{-2} \\[1.0ex] &  \times [1+\log(t/7~ \text{s})]^{-3.02} \Big(\frac{t}{10~\text{s}}\Big)^{-2/3}\\[2ex]& \qquad \qquad \qquad \qquad \qquad \qquad \qquad \quad ,t_\bo<t<t_\s \\[1.5ex]
&~ 9.8\times 10^{5}\text{ K}~  M_{15}^{-3.08} R_{50}^{-1.16} E_{51}^{4.08}\kappa_{0.34}^{3.0} \Big(\frac{t}{10~ \text{m}}\Big)^{-0.83} \\[2ex]&  \qquad \qquad \qquad \qquad \qquad \qquad \qquad  \quad, t_\s<t<t_\tr \\[2ex]
&~ 6.5\times 10^{4}\text{ K}~  M_{15}^{-1.16} R_{50}^{5.0} E_{51}^{1.39}\kappa_{0.34}^{2.23} \Big(\frac{t}{1~ \text{hr}}\Big)^{-5.46} \\[2ex]&  \qquad \qquad \qquad \qquad \qquad \qquad \qquad  \quad, t_\tr<t<t_1 \\[2ex]
&~8.3 \times 10^{4}~ \text{ K}~  M_{15}^{0.05} R_{50}^{0.24} E_{51}^{-0.11}\kappa_{0.34}^{-0.31}\Big(\frac{t}{1~\text{hr}}\Big)^{-0.40} \\[2ex]& \qquad \qquad \qquad \qquad \qquad \qquad \qquad \qquad   t_1<t<t_2 \\[2ex]
\end{cases}
\end{split}
\end{equation}
where
\begin{equation}
t_1 = 57~\text{m} ~M_{15}^{-0.24} R_{50}^{0.94} E_{51}^{0.29} \kappa_{0.34}^{0.50}
\end{equation}
and
\begin{equation}
t_2 = 2.9~\text{hr} ~M_{15}^{-0.77} R_{50}^{0.62} E_{51}^{1.00} \kappa_{0.34}^{1.03} ~.
\end{equation}
%Due to the logarithmic correction, the radiation may reach thermal equilibrium before thet end of the planar phase, depending on the progeitor properties. In that case, the temperature will behave as follows:
%\begin{equation}\label{e:T_obs_eq_BSG}
%\begin{split}
%&T_\cl^{\text{BSG}}(\eta_\lum<1)= \\[1.5ex]
%&\begin{cases}
%&~3.5\times 10^{6} ~\text{ K}~  M_{15}^{-0.26} R_{50}^{0.02} E_{51}^{0.31}\kappa_{34}^{0.01} \Big(\frac{t}{\text{sec}}\Big)^{-0.36}\times\\[1.3ex] &   [1+log(t/t_\bo)]^{-0.01} \qquad \qquad \qquad \qquad \qquad \qquad  , t<t_\s \\[2ex]
%&8.0 \times 10^{4}~ \text{ K}~  M_{15}^{-0.11} R_{50}^{0.37} E_{51}^{0.11}\kappa_{34}^{-0.08}\Big(\frac{t}{\text{hr}}\Big)^{-0.60}  ,t_\s<t
%\end{cases}
%\end{split}
%\end{equation}
At $t_\tr<t$, $\xi_\lum \sim 1$ and Comptonization no longer affects the observed temperature.
In Figure \ref{f:T_obs_BSG} we plot the temperature evolution represented in Eq \eqref{e:T_obs_noneq_BSG}. We also overplot the blackbody temperature of the luminosity shell and the observed temperature that would have been obtained without the logarithmic correction found in this work.
The drop in the observed temperature during the planar phase is significant compared to the uncorrected temperature, and at $t_\s$, $T_\obs$ is an order of magnitude lower than the temperature calculated without the logarithmic correction.
As a result, the decrease in temperature towards thermal equilibrium in the end of the planar phase is much more gradual than previously thought.

%We again point out that between $t_1$ and $t_\tr$ there is a break in $T_\obs$. This break occurs when $\xi=1$, and the contribution from Comptonization, which was significant while the temperature was high compared to $\nu_\min$, disappears.

One can notice that the corrected and uncorrected evolution become identical at $t_\tr\lesssim t$, once the luminosity shell reaches $m_\p$, since the observed radiation now originates in the same shell.

Without the logarithmic correction, the typical energy of the BSG radiation would have exceeded $10$ KeV throughout the planar phase, emitting most of the energy in hard X-rays. 
Nevetheless, the fast decrease in $T_\obs$ brings the typical energy to the extreme UV ($\sim 100$ eV) at the end of the planar phase.
Comparing our results to works that do not assume thermal equilibrium (e.g., \citealt{ensman_borrows92} and NS10), we find higher observed temperatures that exceed their results by about an order of magnitude.
The discrepancy of our results compared to those of NS10 originate mainly from the difference in the normalization of $\eta_\lum$, and from the weaker role of Comptonization imposed by the higher value of $h \nu_\min$, which causes our values of $\xi_\lum$ to be much lower, and result in higher temperatures.

For this BSG progenitor, $R/c\sim 120 \text{s}$. With an effective power law temperaure evolution of $T\propto t^{-1.2}$ at $t<R/c$, light travel time variations will create a spectrum of $F_\nu \propto \nu^{-0.52}$ during the first $\sim 2$ minutes.

%\begin{equation}\label{e:T_obs_eq_BSG}
%\begin{split}
%&T_{\text{obs}}^{\text{BSG}}(\eta_\lum<1)= \\[1.5ex]
%&\begin{cases}
%&8 \times 10^{5} \text{ K}~  M_{15}^{0.05} R_{50}^{0.1} E_{51}^{-0.09}\kappa_{34}^{-0.31} C_1^{-0.05} C_2^{-0.18}  \times\\[1.3ex] & \Big(\frac{t}{t_\bo}\Big)^{-0.33} \qquad \qquad \qquad \qquad  \qquad \qquad \qquad, t<t_\s \\[2ex]
%&2 \times 10^{4} \text{ K}~  M_{15}^{-0.13} R_{50}^{0.33} E_{51}^{0.13}\kappa_{34}^{-0.08} C_1^{0.01} C_2^{0.26}  \times \\[1.3ex] &\Big(\frac{t}{t_{\s,\bo}}\Big)^{-0.57}  \quad \quad \quad \qquad \qquad \qquad \qquad \qquad ,t_\s<t
%\end{cases}
%\end{split}
%\end{equation}

\subsection{Wolf-Rayet}
WR stars are believed to be the progenitors of Ibc SNe. They have lost all of their H envelope (and some of their He) prior to the explosion, and therefore have an electron scattering opacity of $\sim 0.2$ cm$^2$ g$^{-1}$. We use the following properties to describe a typical WR star: $M = 15M_\odot$, $R = 5R_\odot$ and $E = 10^{51} \text{ erg}$. These objects also have radiative envelopes for which $n=3$, and therefore $C_1 = 0.35$ and $C_2 = 0.97$ if the whole envelope is a single polytrope..
For these typical parameters, the breakout time of a WR star would be:
\begin{equation}
t_\bo^{\text{WR}} =  0.12 ~\text{s}~  M_{15}^{0.27} R_{5}^{1.92} E_{51}^{-0.73} \kappa_{0.2} ^{-0.46}~
\end{equation}
and the planar phase ends at:
\begin{equation}
t_\s^{\text{WR}} = 45 \text{ s } M_{15}^{0.42} R_{5}^{1.33} E_{51}^{-0.58} \kappa_{0.2}^{-0.17} ~,
\end{equation}
which, similarly to a BSG, means that the planar phase evolution of a WR explosion currently cannot be resolved in observations.
The luminosity of a typical BSG in the planar and the transition phases is found using equations \eqref{e:L_planar} and \eqref{e:L_transition}:
\begin{equation}
\begin{split}
&L^{\text{WR}}(t) = \\
&\begin{cases}
&2.5\times 10^{45}  \text{ erg}\cdot \text{s}^{-1} M_{15}^{-0.33} R_{5}^{2.31} E_{51}^{0.34} \kappa_{0.2}^{-0.99}\times \\
& [1+ \log(t/0.1~\text{s})]^{-0.01} \Big(\frac{t}{0.1\text{s}}\Big)^{-4/3} \quad \quad ,t<t_\s \\[2ex]
&1.1\times 10^{41}  \text{ erg}\cdot \text{s}^{-1} M_{15}^{-0.74} R_{5}^{0.97} E_{51}^{0.92} \kappa_{0.2}^{-0.82}\times \\[2ex]
& \Big(\frac{t}{1\text{m}}\Big)^{-0.33}  \qquad \qquad \qquad \qquad \qquad ,t_\s<t<t_\tr
\end{cases}
\end{split}
\end{equation}
where $t_\tr=2.7$ minutes.
The small radius of a WR star leads to very high velocities at breakout (almost $\sim 10^{5}$ km s$^{-1}$) resulting in very high temperaures, which makes it hard for the radiation to thermalize. Accordingly, the value of $\eta_\lum$ is very high:
\begin{equation} \label{e:eta_lum_planar_WR}
\begin{split}
\eta_\lum^{\text{WR}}(t) = &343 ~ M_{15}^{-1.59} R_{5}^{-0.77} E_{51}^{2.12}\kappa_{0.2}^{1.66} \Big(\frac{t}{0.1~\text{s}}\Big)^{-1/6} \\ &\times [1+\log(t/0.1 ~\text{s})]^{-1.66} ~,
\end{split}
\end{equation}
while $\xi_\lum(t_\bo) \sim 12$. This means that $\eta_\lum/\xi_\lum\sim 30$, and the radiation is out of thermal equilibrium at breakout. As in the case of BSG explosions, also in WR explosions the radiation remains out of thermal equilibrium throughout the planar phase where $\eta_\lum/\xi_\lum\sim 5$ at $t_\s$.

\begin{figure}
 \centering
\includegraphics[width=1\columnwidth]{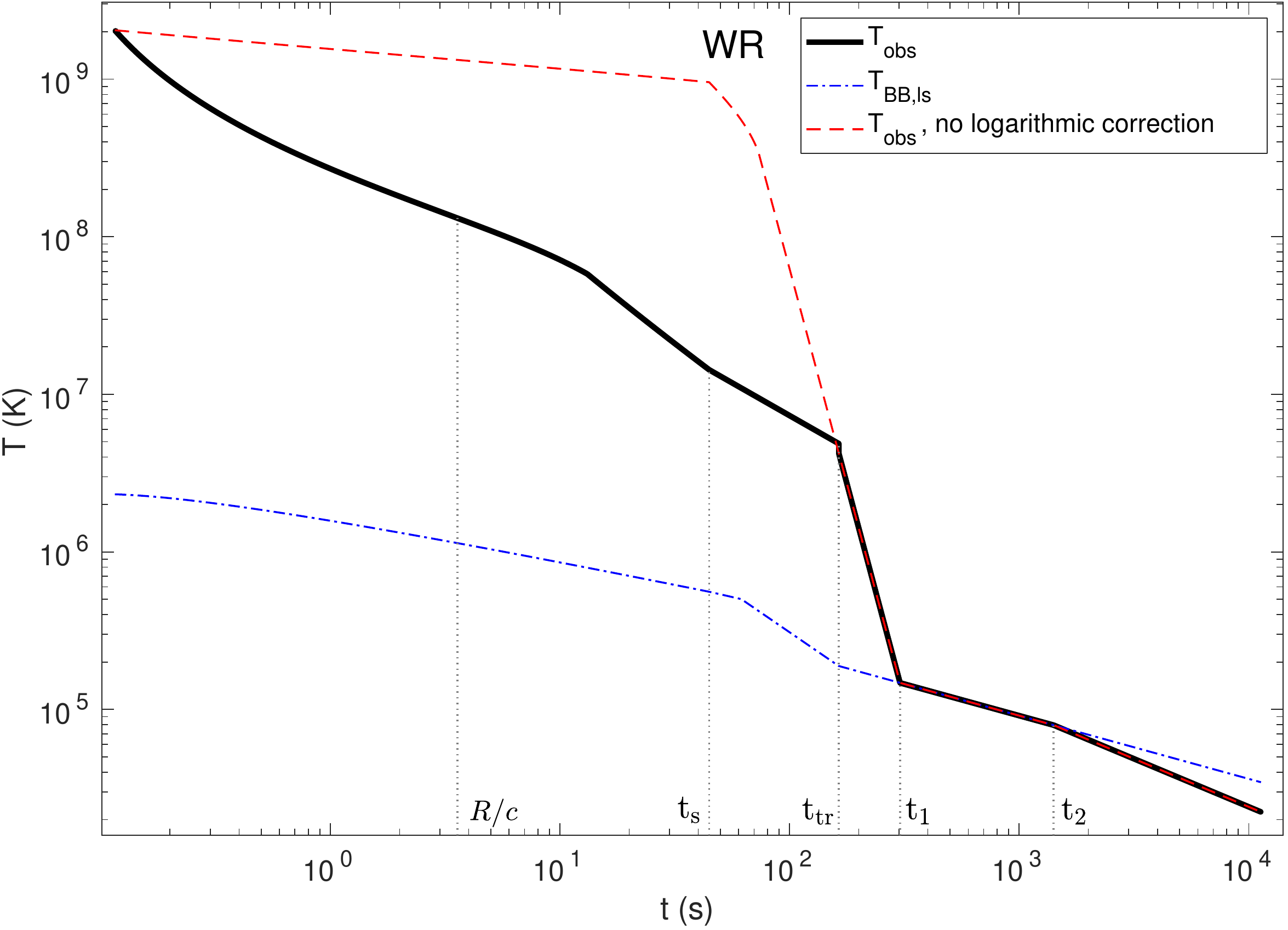} 
\caption{Same as Figure \ref{f:T_obs_RSG} for a WR progenitor with $M_{15}=1$, $R_{5}=1$, $E_{51}=1$ and $\kappa_{0.2}=1$.  The power law values are the same as in Figure \ref{f:T_obs_BSG}. At the end of the planar phase, the radiation peaks in the soft X-rays, instead of in the hard X-ray as would have been expected from the uncorrected temperature. At $t<R/c\sim 12 \text { s}$, light travel time variations create a broad band spectrum in the observer frame, which follows $F_\nu\propto \nu^{-0.52}$.} \label{f:T_obs_WR}
\end{figure}

The observed temperature of the non thermal radiation from a WR explosion, according to equations \eqref{e:T_obs_planar}, \eqref{e:T_obs_sph_noneq_tr_t_t1} and \eqref{e:T_obs_sph_noneq_t1_t_t2} is:
\begin{equation}\label{e:T_obs_noneq_WR}
\begin{split}
&T_\obs^{\text{WR}}(\eta_\lum>1)= \\[1.5ex]
&\begin{cases}
&2.0 \times 10^{9} \text{ K}~  M_{15}^{-3.02} R_{5}^{-1.38} E_{51}^{4.18}  \kappa_{0.2}^{3.02} \\[1.3ex] &  \times \Big(\frac{\xi_\lum}{12}\Big)^{-2} \Big(\frac{t}{0.1 \text{s}}\Big)^{-2/3}[1+\log(t/0.1 ~\text{s})]^{-3.02}\\[2ex]&  \qquad \qquad \qquad \qquad \qquad \qquad \qquad  \quad ,t_\bo<t<t_\s \\[2ex]
&~ 1.1\times 10^{7}\text{ K}~  M_{15}^{-3.08} R_{5}^{-1.16} E_{51}^{4.08}\kappa_{0.2}^{3.0} \Big(\frac{t}{1 \text{m}}\Big)^{-0.83} \\[2ex]&  \qquad \qquad \qquad \qquad \qquad \qquad \qquad  \quad, t_\s<t<t_\tr \\[2ex]
&~ 2.5\times 10^{6}\text{ K}~  M_{15}^{-1.16} R_{5}^{5.0} E_{51}^{1.39}\kappa_{0.2}^{2.23} \Big(\frac{t}{3 \text{m}}\Big)^{-5.46} \\[2ex]&  \qquad \qquad \qquad \qquad \qquad \qquad \qquad  \quad, t_\tr<t<t_1 \\[2ex]
&~1.5 \times 10^{5}~ \text{ K}~  M_{15}^{0.05} R_{5}^{0.24} E_{51}^{-0.11}\kappa_{0.2}^{-0.31}\Big(\frac{t}{5\text{m}}\Big)^{-0.40} \\[2ex]& \qquad \qquad \qquad \qquad \qquad \qquad \qquad \qquad   t_1<t<t_2 \\[2ex]
\end{cases}
\end{split}
\end{equation}
where
\begin{equation}
t_1 = 5~\text{m}~ M_{15}^{-0.24} R_{5}^{0.94} E_{51}^{0.29} \kappa_{0.2}^{0.50}
\end{equation}
and
\begin{equation}
t_2 = 24~\text{m}~ M_{15}^{-0.77} R_{5}^{0.62} E_{51}^{1.00} \kappa_{0.2}^{1.03} ~.
\end{equation}
We plot this temperature evolution in Figure \ref{f:T_obs_WR}.

The breakout temperature of a WR explosion derived using our model exceeds the energy in which pair production starts regulating the temperature ($\sim 50$ keV) and our treatment for the temperature evolution might therefore not be valid in the very early stages of a WR SN.
By the end of the planar phase, the typical energy will have reached $\sim 1\text{keV}$, which lies in the soft X-ray range.

X-rays were associated with the early detecion of the Ib/c SN2008D \citep{Soderberg2008b}, which was discovered during breakout in NGC 2270. However, it is not clear whether the X-ray transient originated in the shock breakout \citep{Soderberg2008b,chevalier_fransson2008} or whether its source was a mildly relativistic jet that penetrated through the envelope of the star \citep [e.g.,][]{Mazzali2008, Xu2008}.

An effective power law fit to the temperature at early times gives $T\propto t^{-0.72}$. Therefore the spectrum at $t<R/c\sim 12 \text{s}$ will follow $F_\nu \propto \nu^{-0.52}$.

\section{Summary}
We derive a self-similar solution for the energy inside a SN envelope when the envelope is fully ionized ($T>1 \text{eV}$) and the SN is in the planar phase ($t<R/v$).
We rely on the exact density and energy profiles of the envelope after shock passage, that serve as initial conditions to the subsequent expansion phase (S60). Our solution is applied to various types of progenitors, and for each we use the appropriate normalization of the density and velocity profiles. Our results for the luminosity and the observed temperature are therefore more accurate than previous analytic solutions that use order of magnitude normalizations.
%However, our models are not corrected for light travel time effects, and are thus applicable only at $t>R_*/c$.
Using a self-similar solution, we find a logarithmic correction to the radiation diffusion in the planar phase, whose strongest effect is on the observed temperature of the SN. We list here our main conclusions:
\begin{enumerate}
\item Since the time doubles itself many times during the planar phase, photons from more internal shells relative to the breakout shell are able to diffuse out through the envelope. The luminosity shell is thus not constant as previously thought, but penetrates farther into the star logarithmically with time. It therefore no longer satisfies $\tau=c/v$. The observed radiation now originates in shells that satisfy $t<t_\diff$. By the end of the planar phase, the mass of the luminosity shell will have increased by a factor of $\sim 10$ relative to the breakout shell.

\item The logarithmic correction to $m_\lum$ highly affects the observed radiation if the luminosity shell is out of thermal euquilibrium during the planar phase. This is more often the situation in compact progenitors with a steep density profile such as BSG and WR stars, where the velocities of the ejecta are very high, and the radiation does not manage to thermalize. The logarithmic correction to the luminosity shell causes the radiation to originate in much denser regions, roughly $10$ times denser than without the corection. This increases the photon production rate by free-free emission by a factor of $\sim 10^2$ and drives the radiation more quickly towards thermal equilibrium.

\item As the luminosity shell recedes into the envelope, the thermal coupling increases with time, causing the observed temperature to decrease more than adiabatically (when the radiation is nonthermal). We reach similar conclusions to previous works, that found the emission from BSG and WR explosions to be non thermal. However, we also show that the typical photon energy is expected to decrease much faster than previously thought, due to the logarithmic correction, which would otherwise cool only adiabatically. The radiation peak is thus directed towards lower frequencies;
a BSG explosion will first peak in the X-rays, but the radiation will reach the soft far UV after $\sim 10$ minutes. A WR star is expected to initially radiate in the hard X-ray to gamma ray regime, and to cool down to the soft X-ray by the end of the planar phase. However, if the model described in this paper yields typical energies that exceed $\sim 50$ KeV, the calculation is no longer valid since the temperature is now regulated by pair production.
%An explosion of a WR star cannot be properly described by our model in the early planar phase since the resulting energies at breakout exceed $\sim 200$ keV. However, the radiation will cool to the soft X-ray by the end of planar phase.

\item Using our accurate self-similar solutions, we find that the radiation from an RSG explosion is expected to be mildly out of thermal equilibrium at breakout, but quickly reach thermalization owing to the logarithmic increase of the luminosity shell. Nevertheless, due to light travel time variations, the observed spectrum will not be that of a blackbody until $t\sim R/c$. Previous works predicted that the radiation from an RSG explosion will be in thermal equilibrium at breakout and throughout the planar phase, despite the fact that these models assumed a constant luminosity shell mass.

\item Due to the weak dependence of the luminosity on the Lagrangian mass coordinate $m$, the logarithmic correction is not expected to have an observed effect on the bolometric luminosity. Owing to that same reason, when the radiation is in thermal equilibrium, the temperature will also not be affected by the logarithmic correction.
\end{enumerate}

\section{Acknowledgments}
This research was supported by an ISF grant and an iCore center. We are grateful to Roni Waldman for providing us with a model of an RSG star and to Luc Dessart and John Hillier for sharing one of their BSG models.

\appendix

\section{Appendix A: The properties of the breakout shell} \label{ap:bo_properties}

\begin{subequations}
\begin{equation}
\begin{split}
v_\bo \footnotemark = 
\begin{cases}
7.1\times 10^{3} \text{ km s}^{-1}  M_{15}^{-0.44} R_{500}^{-0.26} E_{51}^{0.56} \kappa_{0.34}^{0.13}\text{(RSG)}\\\\
2.3\times 10^{4} \text{ km s}^{-1}  M_{15}^{-0.42} R_{50}^{-0.33} E_{51}^{0.58} \kappa_{0.34}^{0.17} , \text{(BSG)} \\\\
4.6\times 10^{4} \text{ km s}^{-1}  M_{15}^{-0.42} R_{5}^{-0.33} E_{51}^{0.58} \kappa_{0.34}^{0.17} , \text{(WR)} ~
\end{cases}
\end{split}
\end{equation}

\begin{equation}
\rho_\bo = 
\begin{cases}
1.2 \times 10^{-9} \text{ g cm}^{-3}  M_{15}^{0.87} R_{500}^{0.51} E_{51}^{-1.13} \kappa_{0.34}^{-1.26}  \text{(RSG)}\\\\
1.6 \times 10^{-10} \text{ g cm}^{-3}  M_{15}^{0.83} R_{50}^{0.66} E_{51}^{-1.17} \kappa_{0.34}^{-1.33}, \text{(BSG)} \\\\
4.5 \times 10^{-10} \text{ g cm}^{-3}  M_{15}^{0.83} R_{5}^{0.66} E_{51}^{-1.17} \kappa_{0.34}^{-1.33} , \text{(WR)}
\end{cases}
\end{equation}

\begin{equation}
u_\bo = 
\begin{cases}
1.6\times 10^{17} \text{ erg g}^{-1} M_{15}^{-0.80} R_{500}^{0.21} E_{51}^{0.87} \kappa_{0.34}^{0.06} , \text{(RSG)}\\\\
1.2\times 10^{18} \text{ erg g}^{-1} M_{15}^{-0.74} R_{50}^{-0.03} E_{51}^{0.92} \kappa_{0.34}^{0.18} , \text{(BSG)} \\\\
4.5\times 10^{19} \text{ erg g}^{-1} M_{15}^{-0.74} R_{5}^{-0.03} E_{51}^{0.92} \kappa_{0.34}^{0.18} , \text{(WR)} ~
\end{cases}
\end{equation}

\begin{equation}
d_{0,\bo} = 
\begin{cases}
6.5 \times 10^{10} \text{ cm} ~M_{15}^{-0.23} R_{500}^{1.90} E_{51}^{-0.23} \kappa_{0.34}^{-0.45}, \text{(RSG)}\\\\
1.7 \times 10^{10} \text{ cm} ~M_{15}^{-0.15} R_{50}^{1.58} E_{51}^{-0.15} \kappa_{0.34}^{-0.29} , \text{(BSG)} \\\\
5.3 \times 10^{8} \text{ cm} ~M_{15}^{-0.15} R_{5}^{1.58} E_{51}^{-0.15} \kappa_{0.34}^{-0.29} , \text{(WR)} ~.
\end{cases}
\end{equation}

\begin{equation}
m_\bo = 
\begin{cases}
2.5\times 10^{29} \text{ g }  M_{15}^{0.44} R_{500}^{2.26} E_{51}^{-0.56} \kappa_{0.34} ^{-1.13} , \text{(RSG)}\\\\
8.7\times 10^{26}  \text{ g }  M_{15}^{0.42} R_{50}^{2.33} E_{51}^{-0.58} \kappa_{0.34} ^{-1.17}, \text{(BSG)} \\\\
7.5\times 10^{24}  \text{ g }  M_{15}^{0.42} R_{5}^{2.33} E_{51}^{-0.58} \kappa_{0.34} ^{-1.17}, \text{(WR)} ~.
\end{cases}
\end{equation}
\end{subequations}
\addtocounter{footnote}{-1}
\footnotetext{Normalized to the breakout velocity before acceleration.}

\section{Appendix B: Glossary of main symbols and notations}
\begin{enumerate}
\item $t$: time

\item $r$: radius

\item $v$: velocity

\item $m$: integrated mass measured from the stellar radius inwards.

\item $\rho$: density

\item $d$: shell width

\item $x$: the distance from the fixed position of the stellar edge at $t=0$. This is also the width of a shell during the planar phase evolution.

\item $u$: energy per unit mass

\item $R_*$: progenitor radius

\item $R_*$: stellar radius

\item $M$: progenitor mass

\item $E$: explosion energy

\item $P$: pressure

\item $L$: observed luminosity

\item $T_\obs$: observed temperature. Defined as the typical photon energy.

\item $\kappa$: opacity

\item $\tau$: optical depth

\item $T_\BB$: blackbody temperature, thermal equilibrium temperature for a given energy density. Defined by Eq \eqref{e:T_BB}.

\item Breakout shell: the shell from which the shock breaks out. Satisfies $\tau=c/v$ during the planar phase.

\item Luminosity shell: the shell that is the source of the observed luminosity. 

\item Colour shell: the shell where the observed temperature is determined. Coincides with the shell satisfying $\eta=1$ when the radiation is in thermal equilibrium and with the luminosity shell when the radiation is out of thermal equilibrium.

\item For any quantity $x$, we use the following subscripts:

$x_\bo$: value at the breakout shell at breakout.

$x_\lum$: value at the luminosity shell.

$x_\cl$: value at the colour shell.

\item $m_\bo$: the mass of the breakout shell.

\item $m_\pl$: the mass of the luminosity shell at the end of the planar phase (at $t=t_\s$).

\item $t_\bo$: time of breakout

\item $t_\diff$: diffustion time

\item $t_\s$: time of breakout shell transition to the spherical phase

\item $t_{\text{spherical}}(m)$: the time a shell of mass $m$ enters the spherical phase. Defined in Eq \eqref{e:t_s}.

\item $t_\eq$: the time when the radiation reaches thermal equilibrium during the planar phase, in case it was out of thermal equilibrium at breakout. 

\item $t_\tr$: time when the transition phase ends.

\item $t_\c$: the first time the shell that satisfies $\tau = c/v$ during the transition phase reaches a shell in which the radiation is in thermal equilibrium. Thermal equilibrium in that shell was achieved by photons produced at the end of the planar phase.

\item $t_{2,\d}$: the first time the shell that satisfies $\tau=c/v$ coincides with the shell satisfying $\eta=1$ during the transition phase.

\item $t_1$: the first time in which the observed radiation is in thermal equilibrium, in case it was not in thermal equilibrium at the end of the transition phase. Thermal equilibrium was achieved by photons produced at an earlier time.

\item $t_2$: the first time in which the radiation is in thermal equilibrium achieved by photons produced at $t_2$, in case it was not in thermal equilibrium at the end of the transition phase.

\item $\eta$: thermal coupling coefficient. Defined in Eq \eqref{e:eta_def}.

\item $\xi$: logarithmic Comptonization factor. Defined in Eq \eqref{e:xi_def}.

\item $n$: power law index describing the pre-explosion stellar density profile. $n$ is related to the adiabatic index $\gamma$ by $\gamma = 1+1/n$.
\end{enumerate}
\vspace{1cm}
\bibliographystyle{apj}	
\bibliography{self_similar_paper_ApJ}
\end{document}